\documentclass[aps,pra,twocolumn,groupedaddress,showpacs,showkeys,preprintnumbers]{revtex4}
\usepackage{graphics}
\usepackage{amsmath}

\DeclareMathOperator{\rot}{rot}
\def\SCDOT{\hspace{-0.8mm} \cdot \hspace{-0.8mm}}

\begin{document}


\title{Alternative to the principle of constant speed of light\footnote{
A history of revisions is given in section \ref{SSEC:REVISION}}}


\author{Herbert Wei\ss}
\email[\emph{Email: }]{hw@terminal-e.de}
\affiliation{Am Bucklberg 19, D-83620 Feldkirchen-Westerham, Germany}


\date{\today}

\begin{abstract}
To consider a medium carrying light and electromagnetic waves is impossible, when this medium shall have properties according to the principle of constant speed of light, that is, isotropy of speed of light in every system of reference. Therefore, with principle of constant speed of light abandoned, the so-called \emph{medium synchronization} of clocks is defined, yielding isotropy solely in the system at rest and anisotropy in all moving systems. From medium synchronization the appropriate coordinate transformation is developed, the so-called \emph{medium transformation}, a Galilean type of transformation, thus maintaining simultaneity between all participant systems of reference and changing the form of Maxwells equations to the extended Maxwell equations. Despite the fact that medium transformation violates Lorentz invariance and the principle of relativity, it is demonstrated that the results from medium transformation and extended Maxwell equations are fully compliant with observable phenomena. Moreover, it is shown that the concept of a medium, a preferred frame of reference, is compatible not only with medium transformation, but also with Lorentz transformation and with theory of special relativity, so it is quite possible now to consider a realistic medium. Finally, a discrepancy is exposed within theory of special relativity, giving rise to assume a preferred frame of reference even in special relativity.
\end{abstract}


\maketitle



\section{Introduction}

After prediction of electromagnetic waves by J.\ C.\ Maxwell and their demonstration by Heinrich Hertz it has been obvious to assume a medium
\footnote{
The commonly used name for the carrier of light and electromagnetic waves is, of course, `aether' or `ether'. There are numerous, mostly visionary, hypotheses on this aether assuming numerous, partly rather different properties. Here, we talk about one robust property, which can be found in most of all other carrier of waves: the anisotropy of speed of waves in moving frames of reference. This property is common to only a few aether models. Because we treat the aether as a realistic carrier, and in order to distinguish our model from other models of aether, we denote the carrier of light and electromagnetic waves as `medium'.}
  carrying those waves. Many attempts have been made on measuring a drift velocity of the earth with respect to this medium, but all of them failed. Albert Einstein solved this problem with the \emph{principle of constant speed of light}, an axiom leading to Lorentz transformation and to the well-known and very successful theory of special relativity \cite{Einst05}.

The principle of constant speed of light demands isotropy of speed of light in every inertial system, which is unfortunately incompatible with the properties of a realistic medium, thus a medium moved out of the focus of science and no more serious research has been done on that subject. But anyway, two charged bodies or simply two massive bodies some distance apart exert forces to each other, forces, which must be carried by something concrete. Light and also particles have properties of waves that give rise to ask for the oscillating matter. On one hand only a medium can explain the nature of forces and electromagnetic waves, on the other hand general isotropy of speed of light makes it impossible to consider a realistic medium. One way to get out of this dilemma is to change the demand for general isotropy of speed of light, so we have to ask: is the principle of constant speed of light the only possible presupposition to attain a transformation of coordinates corresponding with observable phenomena or is there an alternative given compatible with the concept of a realistic medium?

\subsection{The problem\label{SSEC:PROBLEM}}

The answer to the question above is: {\scshape yes}! Indeed, there is an alternative! But this is not the problem, anisotropy of speed of light assumed for moving frames of reference leads in a straightforward manner to a transformation of coordinates complying with a medium. The problem arises from Maxwell equations. -- Since light is a kind of an electromagnetic wave, Maxwell equations must be taken into account, that is, the selected transformation of coordinates must yield reasonable results when applied on Maxwell equations, the results must comply with observable phenomena, otherwise this transformation will not be considered at all, it will be rejected with good reasons. -- This is the first part of the problem.

On one hand, Lorentz transformation is the only transformation known yielding reasonable results when applied on Maxwell equations. But Lorentz transformation, derived from the principle of constant speed of light \cite{Einst05}, is incompatible with the concept of a medium, a preferred frame of reference. -- On the other hand, transformations complying with the concept of a realistic medium are of Galilean type, because they maintain simultaneity between all participant systems of reference. Unfortunately, all these transformations change the form of Maxwell equations, hence the results from the transformed equations haven't been expected to comply with observable phenomena. The commonly accepted principle of relativity means that the laws of physics must maintain their form under a transformation of coordinates. Mainly, this dogma is the obstacle for a discussion of Galilean transformations. There are no reasons known to consider and to suspend principle of relativity, for example, a discrepancy within the framework of special relativity. -- This is the second part of the problem.

\subsection{Attempts to solve the problem}

Any solution to our problem must involve anisotropy of speed of light and the concept of a preferred frame of reference. Einstein-aether theories, like that of Eling et.\ al.\ \cite{Eling:gr-qc/0410001v2}, introduce terms of breaking Lorentz symmetry. All those attempts to establish the idea of a preferred frame of reference using a Riemannian or a post-Riemannian space-time need not being considered, because speed of light in those equations is assumed constant and is mostly set to unity. Anisotropy of speed of light isn't represented adequately in those theories, but anisotropy of speed of light is the one and only property of our moving frames of reference, therefore these articles cannot contribute to our problem. From the remaining articles dealing with Galilean transformations, the article from Puccini \cite{Puccini:physics/0407096v1} is outstanding and interesting \footnote{
I got notice of Puccini's article just after publication of version 8. Due to the relevance I changed the article accordingly.}. This article gives an overview of recent working on the problem. Puccini employs the same transformation used in this paper, the medium transformation, and derives the same extended Maxwell equations as given in section \ref{SEC:MAXWELL}, including the inhomogenious version. He explains satisfactorily the Wilson effect, an experiment actually intended to prove theory of special relativity, so this is quite a step on the way to the solution, but this result was obviously insufficient to initiate the discussion of principle of relativity.

\subsection{Solution of the problem}

As mentioned above, the solution of our problem must come from the class of Galilean transformations. These transformations are not compatible with the principle of constant speed of light. Moreover, a preferred frame of reference is also incompatible with the principle of relativity. Thus both principles must be considered, and principle of constant speed of light must be abandoned. This is a big affair, so we have to work out very carefully the basics for the proposed transformation of coordinates, in order to provide a well founded, solid basis for the derivation of medium transformation.

The measurable properties of light propagation, including anisotropy, are governed by the synchronization scheme applied to the clocks of the reference system. For instance, isotropy of speed of light is realized by applying Einstein synchronization on the clocks of the reference system. It is not a naturally given phenomenon but a manually determined. The synchronization scheme applied to the clocks is responsible for observable properties of light propagation, nothing else. Therefore, we have a look more closer on the method of synchronization.
Beforehand, in section \ref{SEC:DEFS}, we give the definition of the terms `reference system', `inertial system', and `coordinate system'.
Then, the methods of synchronization are discussed in section \ref{SEC:SYNCHRONIZATION}. We ask for the conditions to be met by the synchronization method to comply with observable phenomena. Subsequently, we define a synchronization scheme, the medium synchronization, applicable for different systems of reference \emph{and} compatible with the properties of a realistic medium. In section \ref{SEC:MT} we develope the transformation of coordinates complying with medium synchronization, which we call medium transformation. The relation between medium transformation and Lorentz transformation is given in section \ref{SEC:MT:LT}.

With this thoroughly developed coordinate transformation in hand, we focus on the problem proper. In section \ref{SEC:MAXWELL} and appendixes \ref{APP:MT:MAX:DC} and \ref{APP:DER:WAVE:EQ} we demonstrate that a coordinate transformation of Galilean type, the medium transformation, can be successfully applied on Maxwell equations, leading to a new form of extended Maxwell equations, namely Eqs. (\ref{MAX.DC.MT}) and (\ref{MAX.IN.MT}). The extended Maxwell equations have reference to the preferred frame with a term featuring the drift velocity \(\bf v\) of the moving frame, thus compatible with the concept of a priviledged frame. But this is only half the way, because the new Maxwell equations of its own primarily say nothing. We have to show meaningful applications in order to verify medium transformation and extended Maxwell equations. To do so, we derive the wave equation (\ref{WAVE.EQ.MT}) from the extended Maxwell equations, and we transform the wave function. Then we plug in the medium transformed wave function (\ref{WAVE.FUNC.MT}) into the extended wave equation in order to show that in the moving frame \(K'\) the wave equation recommends the same observable properties from a wave function, condition (\ref{CPP.EP.ZERO}), as the wave equation for a frame at rest, condition (\ref{CP.E.ZERO}). So far, the first part of our problem is solved.

The results of sections \ref{SEC:DEFS} to \ref{SEC:MAXWELL} are summarized in section \ref{SEC:CONCLUSION}, and we ascertain the compliance of Lorentz transformation and theory of special relativity with the concept of a medium, a preferred frame of reference. In appendix \ref{APP:MOV:CLKS}, where the transport of time with moving clocks is discussed, we expose an intrinsic contradiction of special relativity. Theory of special relativity admits any number of \emph{different} results at the same time, but only one single result can be expected from the experiment. This problem can be solved only when the principle of relativity becomes suspended at least for the scope covering this experiment. Therefore, a discussion of principle of relativity is mandatory, and consequently a discussion of Galilean transformations is possible, thus the second part of our problem is solved as well.

We shall mention, that a restriction of principle of relativity is a problem at most to theory of special relativity and perhaps to theories based on special relativity, but it is not a problem to physics.

\subsection{Remarks on the medium}

The purpose of this paper is, in fact, to yield the presupposition to consider a realistic medium, but the medium itself isn't discussed at all. This might be unsatisfactory to readers, who want come to know something about the medium. Therefore I added appendix \ref{APP.MED} giving a brief description of the current ideas to a model of the medium. But I have to underline, it is far from being a theory of the medium, appendix \ref{APP.MED} is only a concise summary of ideas.

\subsection{Formalism and system of units\label{SSEC:FORMALISM}}

\begin{figure}
	\centering
		\includegraphics{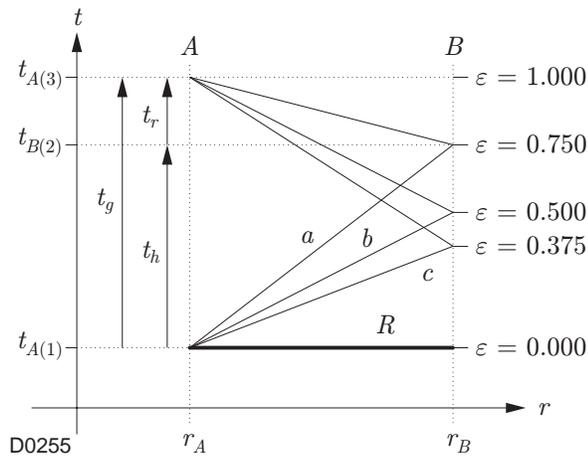}
	\caption{Synchronization parameter \(\varepsilon\). The bold horizontal line \(R\) represents a rod of length \(l=r_B-r_A\) resting in the stationary system \(K_0\) at instant \(t_{A(1)}\). The \(r\)-coordinate is common to the axis of the rod given by \(r={\bf r}/\widehat{\bf r}\). The dotted lines \(A\) and \(B\) are the world lines of endpoints \(A\) and \(B\) of the rod. Lines \(a\), \(b\), and \(c\) represent three possible world lines of a light pulse emitted from \(A\), traveling to \(B\) and back to \(A\) after reflection at \(B\). The instant of reflection \(t_{B(2)}\) can have any value between emission at \(t_{A(1)}\) and absorption at \(t_{A(3)}\). Due to Eq.~(\ref{EPS.DEF}) the corresponding synchronization parameter \(\varepsilon\) must range from 0 to 1.}
\label{fig:d0255}
\end{figure}

All equations are shown exclusively in SI units using three-vectors with time as a regular parameter. In fact, there is a reason for treating time different from position coordinates, because time is strictly monotone, the direction of flow is never reversed in contrast, for example, to the movement of a mechanical oscillator. Regarding any process a particular position coordinate can occure more than once, a particular time coordinate only once. We don't use four-dimensional notation with Lorentz invariant Minkowsky spacetime, because medium transformation violates Lorentz invariance (and I have to admit being not sure whether Minkowsky spacetime applies in this case). Time coordinates will be denoted by uppercase indices (\(t_A\), \(t_B\), \ldots), time differences, i.e., elapsed times, by lowercase indices (\(t_g\), \(t_h\), \(t_r\), \ldots). Figures \ref{fig:d0255}, \ref{fig:d0257}, \ref{fig:d0258}, \mbox{and \ref{fig:d0262}} are showing space-time-like coordinate systems. World lines of locations are represented by thin dotted lines, locations pertaining to the system at rest as well as events are denoted by uppercase letters (\(A\), \(B\), \mbox{\(C\), \ldots)}, and locations pertaining to the moving system are denoted by primed uppercase letters (\(A'\), \(B'\), \mbox{\(C'\), \ldots)}.

In order to enhance the clarity and legibility of equations, we use a parenthetical notation (\ref{SCAL:PROD:PAR}) for the scalar product besides the notation with the center dot.
\begin{equation}
  \nabla \cdot {\bf E}
   								= \left( \nabla {\bf E} \right)
	 								= \frac {\partial E_x}{\partial x}
									+ \frac {\partial E_y}{\partial y}
									+ \frac {\partial E_z}{\partial z}
	\label{SCAL:PROD:PAR}									
\end{equation}

The differential operators \(\nabla\), \(\Delta\), and \(\rot\), applied on primed magnitudes \({\bf E}'\) and \({\bf B}'\) shall adapt correspondingly, \mbox{Eqs.\ (\ref{NABLA:PRIME}--\ref{ROT:PRIME})}.
\begin{equation}
	\nabla \cdot {\bf E}' = \left(\begin{array}{c}
										\frac {\partial}{\partial x'}	\vspace{0.5em} \\
									  \frac {\partial}{\partial y'} \vspace{0.5em} \\
										\frac {\partial}{\partial z'}
										\end{array}\right)
										\cdot {\bf E}'
	\label{NABLA:PRIME}
\end{equation}

\begin{equation}
	\Delta {\bf E}' = \left(
											\frac{\partial^2}{\partial x'^2}
									  + \frac{\partial^2}{\partial y'^2}
										+ \frac{\partial^2}{\partial z'^2}
										\right)
										{\bf E}'
	\label{DELTA:PRIME}
\end{equation}

\begin{equation}
	\rot {\bf E}'= \nabla \times {\bf E}' = \left( \begin{array}{c}
											\frac{\partial E'_z}{\partial y'} - \frac{\partial E'_y}{\partial z'}
											\vspace{0.5em} \\
											\frac{\partial E'_x}{\partial z'} - \frac{\partial E'_z}{\partial x'}
											\vspace{0.5em} \\
											\frac{\partial E'_y}{\partial x'} - \frac{\partial E'_x}{\partial y'} 
										\end{array} \right)
	\label{ROT:PRIME}
\end{equation}

Similarly, dots on primed symbols denote derivatives of primed time \(t'\), i.e.,
\begin{equation}
	\ddot{\bf E}'=\frac{\partial^2}{\partial t'^2}{\bf E}'.
\end{equation}

It is frequently switched between vectorial representation in boldface symbols, like \(\bf E\), and representation in component form, like \(\left(E_x, E_y, E_z \right)\). When the component form is used, the vector of drift velocity \(\bf v\) is assumed to have only one component in the \(x\)-direction according to Eq.~(\ref{DRIFT:VECTOR}), but we do not mention this at every instance the form is changed. We kindly ask to indulge this simplification.

\subsection{History of revisions\label{SSEC:REVISION}}

{\footnotesize
\noindent {\bf v2}\ldots{\bf v5}: typos in equations corrected -- {\bf v6}: differential operators expounded in subsection \ref{SSEC:FORMALISM} -- app.\ \ref{APP:MOV:CLKS} enhancement added -- minor textual changes -- {\bf v7}: matrix notation introduced -- medium transformation expounded in detail -- use of synchronization parameter clarified -- lenght contraction and time dilation added -- grammatical and textual changes -- {\bf v8}: textual changes -- grammatical errors corrected -- relation of synchronization parameter \(\varepsilon\) to the ratio \(v_{h}/v_{r}\) established in amendment to subsection \ref{SSEC:SYNCMETHLIGHT} -- major changes in subsections \ref{SSEC:WAVE:REST} and \ref{SSEC:WAVE:MOVING} aiming on the developement and transformation of the wave vector \(\bf E\) to become as clear as possible. -- {\bf v9}: problem and solution clearly emphasized in the introduction -- Puccini's contribution \cite{Puccini:physics/0407096v1} included -- abstract and conclusions changed.
}

\section{Reference system, coordinate system and time coordinate\label{SEC:DEFS}}

Throughout this article we are dealing with coordinates and coordinate transformations. Every set of coordinates pertains to a coordinate system, and this coordinate system in turn pertains to a reference system. In order to clarify the usage and the meaning of these terms, and to avoid misunderstandings, we give the definitions according to Mittelstaedt \cite{Mittelstaedt}. Furthermore, the definition of an inertial system leads in a natural way to the synchronization procedure for the clocks, or more precisely, it yields the requirements for the synchronization procedure. And here, I'll point out, the time coordinate, represented by the clocks of a system and the applied synchronization scheme, is the central subject of this paper.

\subsection{Reference system and inertial system}

A reference system \(S\) in our sense is a rigid material basis equipped with clocks and measuring-rods. A special kind of a reference system is an \mbox{inertial system \(I\).}
\vspace*{-2.8mm} \\
\fbox{
\begin{minipage}[c]{81.6mm} 
An \emph{inertial system} is a space-time-like reference system, where noninteracting mass points \(k_{1}, k_{2}, \ldots k_{n}\) are moving uniformly on straight lines \cite[p.~35]{Mittelstaedt}.
\end{minipage}
} 
\vspace*{-0.8mm} \\ 
Noninteracting means: no forces act on the mass points.

\subsection{Coordinate system}

By means of a coordinate system \(K\) every point \(P\) of space a position vector \(\bf r\) is assigned (see \ref{POS.COORD}). Together with time \(t\), given by the local clocks, a space-time-like coordinate system \(K\left({\bf r}, t\right)\) is defined. Here, we use only coordinate systems floating along with the reference system \(I\) (the material basis), i.e., coordinate sytems, where position coordinates \({\bf r}\left(P_I\right)\) of locations \(P_I\) pertaining to the reference system \(I\) are constant, independent of time: \mbox{\(\partial {\bf r}\left(P_I\right)/\partial t=0\)}.---As an example, the optical bench in a lab is a system of reference and the coordinate system may be established by rulers mounted on the optical bench.---Furthermore we restrict on right-basis orthogonal coordinate systems. For short we denote such a coordinate system as frame, i.e., the right-basis orthogonal coordinate system at rest will be denoted as the frame at rest.

\subsubsection{Position coordinates}

\begin{figure}
	\centering
		\includegraphics{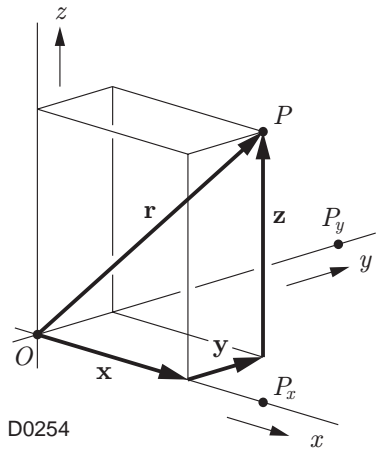}
		\caption{Right-basis orthogonal coordinate system \(K_0\) and position vector \(\bf r\) with components \(\bf x\), \(\bf y\), and \(\bf z\). Origin \(O\) and points \(P_x\) and \(P_y\) give the direction of the \(x\)-, and \(y\)-axis. The \(z\)-axis is determined by the requirements for a right-basis orthogonal coordinate system.}
	\label{fig:d0254}
\end{figure}

\label{POS.COORD} We need at least three points to determine the three axes of coordinate system \(K_0\), first the origin \(O\) of our coordinate system, second a point \(P_{x}\) giving the direction of the \(x\)-axis by line \(\overline{OP_x}\), and third a point \(P_{y}\) giving the direction of the \(y\)-axis by line \(\overline{OP_y}\), whereas point \(P_{y}\) has to be chosen in such a manner that the \(y\)-axis is perpendicular to the \(x\)-axis (Fig.~\ref{fig:d0254}). The direction of the \(z\)-axis is determined by the \(x\)- and \(y\)-axis due to the requirements for a right-basis orthogonal coordinate system. The position vector \(\bf r\) is the sum of vectors \(\bf x\), \(\bf y\), and \(\bf z\), Eq.~(\ref{POS.VEC.DEF}). Vector \(\bf x\) is aligned with the \(x\)-axis, vector \(\bf y\) is aligned with the \(y\)-axis, and vector \(\bf z\) is aligned with the \(z\)-axis. 
\begin{equation}
	{\bf r}= {\bf x}+{\bf y}+{\bf z}
	\label{POS.VEC.DEF}
\end{equation}
The vectors \(\bf x\), \(\bf y\), and \(\bf z\) are given by Eqs.~(\ref{XYZ.VEC})
\begin{equation}
	  {\bf x}= x\, \widehat{{\bf x}}, \hspace{10mm}
	  {\bf y}= y\, \widehat{{\bf y}}, \hspace{10mm}
	  {\bf z}= z\, \widehat{{\bf z}}
	\label{XYZ.VEC}
\end{equation}
with the unit vectors \(\widehat{\bf x}\), \(\widehat{\bf y}\), and \(\widehat{\bf z}\) \footnote{
Usually, the unit vectors comprising the basis of a right-handed, orthogonal, coordinate system are denoted by \(\bf i\), \(\bf j\), and  \(\bf k\). Because we use the symbol \(\bf k\) for the wave vector, we treat the unit vectors \(\widehat{\bf x}\), \(\widehat{\bf y}\), and \(\widehat{\bf z}\) as the basis of coordinate system \(K_0\) in order to avoid misunderstandings.}, 
each parallel to the corresponding axis. By means of a measuring-rod \(R_m\) of unit lenght we are now able to measure the three quantities \(x\), \(y\), and \(z\), usually called the \emph{coordinates} of position vector \(\bf r\). The measuring-rod also defines the length of unit vectors \(\widehat{\bf x}\), \(\widehat{\bf y}\), and \(\widehat{\bf z}\). For short, vector \(\bf r\) is commonly denoted only by the coordinates \(x\), \(y\), and \(z\) in parentheses (\ref{POS.VEC.R}) with the unit vectors \(\widehat{\bf x}\), \(\widehat{\bf y}\), and \(\widehat{\bf z}\) being omitted.
\begin{equation}
	{\bf r} = \left(\begin{array}{c}
										x\\y\\z
									\end{array}\right)
	\label{POS.VEC.R}
\end{equation}

\subsubsection{Coordinate transformation and matrix notation\label{SSS:COOTRANS}}

As soon as we are dealing with more than one reference system, e.g., coordinate system, we encounter the problem of relating the coordinates of the participant coordinate systems. In order to distinguish magnitudes from different coordinate systems, we denote magnitudes of the system at rest \(K_0\) by regular, unprimed symbols (\({\bf r}, x, y, z, t\)), and we denote magnitudes of other coordinate systems \(K'\), especially of moving systems, by primed symbols (\({\bf r}', x', y', z', t'\)). Position vector \(\bf r\) from Eq.~(\ref{POS.VEC.R}) identifies point \(P\) with respect to coordinate system \(K_0\). The \emph{same} point of space is identified by vector \({\bf r}'\) pertaining to coordinate system \(K'\) given by Eq.~(\ref{POS.VEC.R.PR}). See also Fig.~\ref{fig:d0261}. Measuring-rod \(R'_m\) used in coordinate system \(K'\) and measuring-rod \(R_m\) are assumed to have equal properties. Exchanging the measuring-rods must not affect the values of the coordinates measured.
\begin{equation}
	{\bf r}' = \left(\begin{array}{c}
										x'\\y'\\z'
										\end{array}\right)
	\label{POS.VEC.R.PR}
\end{equation}

Coordinates \(x'\), \(y'\), and \(z'\) of vector \({\bf r}'\) are calculated from coordinates \(x\), \(y\), and \(z\) of vector \(\bf r\) by applying the appropriate coordinate transformation. When the position of point \(P\) depends on time in any of the two involved coordinate systems, the time coordinates also have to be taken into account. Thus the general form of a \emph{coordinate transformation} gives the relation of coordinates \(x'\), \(y'\), \(z'\), and \(t'\) of the (moving) system \(K'\) to the coordinates \(x\), \(y\), \(z\), and \(t\) of the system (at rest) \(K_0\).
A convenient notation for this relation is the matrix equation (\ref{MATRIX:EQ:TR}), where the time coordinates \(t\) and \(t'\) are simply added to the position vectors \({\bf r}\) and \({\bf r}'\), respectively.
\begin{equation}
	\left(\begin{array}{c} c_0 t'\\{\bf r}'\end{array}\right)
	= M \left(\begin{array}{c} c_0 t\\{\bf r} \end{array}\right)
	\label{MATRIX:EQ:TR}
\end{equation}

In order to make the time coordinates space-like, the time coordinates \(t\) and \(t'\) have been multiplied by the vacuum speed of light \(c_0\). This is actually not necessary, but with this manipulation the elements of the matrix for Lorentz transformation (\ref{MATRIX:EQ:LORENTZ}) directly show the symmetry properties of the transformation.
Equation (\ref{MATRIX:EQ:TR}), written completely in components, reads
\begin{equation}
	\left(\begin{array}{c} c_0 t'\\x'\\y'\\z'\end{array}\right)
	= \left(\begin{array}{cccc} m_{00} & m_{01} & m_{02} & m_{03} \\
														  m_{10} & m_{11} & m_{12} & m_{13} \\
														  m_{20} & m_{21} & m_{22} & m_{23} \\
														  m_{30} & m_{31} & m_{32} & m_{33} 
					\end{array}\right)
	  \left(\begin{array}{c} c_0 t\\x\\y\\z \end{array}\right).
	\label{MATRIX:EQ:TXYZ}
\end{equation}

The definition of matrix \(M\) is easily found by comparing Eqs.~(\ref{MATRIX:EQ:TR}) and (\ref{MATRIX:EQ:TXYZ}) and is given here for completeness:
\begin{equation}
	M
	= \left(\begin{array}{cccc} m_{00} & m_{01} & m_{02} & m_{03} \\
														  m_{10} & m_{11} & m_{12} & m_{13} \\
														  m_{20} & m_{21} & m_{22} & m_{23} \\
														  m_{30} & m_{31} & m_{32} & m_{33} 
					\end{array}\right).
	\label{MATRIX}
\end{equation}

Matrix \(M\) represents the coordinate transformation from system \(K_0\) into system \(K'\).
We have to find the elements \(m_{ik}\) of matrix \(M\) so that observables, measured in both systems \(K_0\) and \(K'\), show properties complying with a realistic medium.

\subsubsection{Time coordinate}

The time coordinate \(t\) of our reference system is represented by the clocks and their zero position settings. We have to choose the time coordinate in such a manner, that noninteracting mass points appear to move uniformly, that is, when the time-path diagrams of all those mass points become straight lines. We assume, all our clocks are of identical construction but initially with randomly distributed zero position settings. To achieve a reasonable common time coordinate for our reference system, we have to \emph{adjust} the zero position settings of all clocks in order to be able to \emph{measure} the constant velocity of the above mentioned uniformly moving mass points. This kind of adjustment the zero positions we call \emph{synchronization of clocks}.

\section{Synchronization of clocks\label{SEC:SYNCHRONIZATION}}

To synchronize the clocks of our reference system we have to select one particular clock, for instance, the clock at the origin \(O\) of the coordinate system, to be the reference clock. The distances of all other clocks to this reference clock are given by the position coordinates of the clocks. Then we need a facility moving uniformly at constant velocity \(v_h\) along a straight line crossing the origin as well as the position of the clock to be synchronized. Both clocks store the time at the instant the facility is passing. When our facility has passed the second clock, the movement is reversed and our facility drifts back at constant velocity \(v_r\) on the same path passing both clocks in opposite order. Again, both clocks store its time at the instant the facility is passing. By means of the given ratio \(v_{h}/v_{r}\), the positions of the clocks, and the locally recorded time coordinates we calculate the velocities \(v_{h}\) and \(v_{r}\) and furthermore the offset for synchronization of the clock. This procedure has to be repeated for all clocks in our inertial system.

Another synchronization method appearing in literature is done by slowly transport of clocks \cite[p.~115]{Mittelstaedt}. The problem with this method is that the clocks have to move infinitely slowly, which is not practicable. At finite velocities the clocks no more transport pure Lorentzian time, therefore, this kind of synchronization will not be discussed here. Lorentzian time is the time given by the clocks in an Einstein synchronized inertial system \(K_0\). A clock is said to transport Lorentzian time when the reading of the moving clock is always the same as the reading of a stationary clock of system \(K_0\) at the instant the two clocks meet. For some details about the transport of clocks at real speed see appendix \ref{APP:MOV:CLKS}.

Our facility can be a massive body as well as a signal traveling as wave. The most commonly used facility for synchronization of clocks are light signals. No matter, which facility is used, we have always to know exactly the ratio \(v_{h}/v_{r}\), otherwise our synchronization result will be ambiguous. We mention this point, because with light signals we just have this situation, the ratio \(v_{h}/v_{r}\) is not known because the speed of the `carrier' of the signals with respect to our system of reference cannot be measured (see below). To overcome this problem, we have to introduce something like a rule giving us the ratio \(v_{h}/v_{r}\).
Before we are talking about such rules, we will have a look at the results concerning the experiments for measuring a drift velocity of the earth (our system of reference) with respect to a medium (our carrier), and we mention briefly the results of measurement of speed of light. Then we describe the synchronization procedure with light signals in detail and show two applicable `rules', i.e., schemes of synchronization: Einstein synchronization and medium synchronization.

\subsection{Measurement of drift velocity}

After discovery of electromagnetic waves and the wave properties of light, there have been a lot of attempts on measuring the drift velocity of the earth with respect to a medium. The most popular of all was that of Michelson and Morley \cite{Michelson}, but all experiments had the same result: there was no drift velocity measurable, independent from the orientation of the legs of the interferometer.

There are several hypotheses to explain this result, but only two are of interest: --i-- Albert Einstein assumed that speed of light in vacuum has generally the same magnitude, regardless to the direction and regardless to the moving state of the observer. He couldn't explain this behaviour, and, guided by the experimental results, he treated his assumption as an axiom and rised it to the principle of constant speed of light. Einstein denied to deal with a medium, a luminiferous aether, because there is no velocity vector known to be assigned to a point of space, where electromagnetic processes take place, he said. And in fact, the principle of constant speed of light leads to a consistent theory of coordinate transformation, the well-known theory of special relativity. --ii-- H.\ A.\ Lorentz (and formerly FitzGerald) assumed the legs of the interferometer in the drift direction being shortened by an amount, which compensates for the longer propagation time in that leg. Lorentz successfully employed this assumption to describe electromagnetic phenomena in moving systems \cite{Lorentz04}. He concluded, a drift velocity of any magnitude cannot be measured by the method of Michelson and Morley (and all equivalent methods). This explanation sounds slightly strange, but Lorentz transformation as derived and used in theory of special relativity just reveals such a behaviour of our material basis.

At this time principle of constant speed of light is the only explanation for the results of the Michelson and Morley experiment, and we have to admit that the results prefer this kind of explanation. But we will show that another explanation is possible and reasonable, an explanation giving properties to space similar to those of a medium and leading to a behaviour of bodies just as supposed by Lorentz. Despite the fact that Michelson and Morley measured a drift velocity of zero at all possible conditions, we suppose a drift velocity is not measurable by this method, so we assume the drift velocity being still unknown.

\subsection{Measurement of speed of light}

The experiments designed for measuring speed of light use a closed loop path, so that the propagation time of the light signal can be measured by a single clock. Thus we can only measure the mean of speed of light. For empty space this mean value is denoted by \(c_0\) and is given by the sum of all distances \(s_i\) the light signal propagates, divided by the sum of all propagation times \(t_i\) of that closed loop path (\ref{C.NULL}).
\begin{equation}
	\frac{\sum s_i}{\sum t_i}=c_0
\label{C.NULL}
\end{equation}

Although the experiments differ in the number of paths and the directions, the mean of speed of light in empty space has always the same value of \(c_0\), regardless to the orientation and to the state of movement the apparatus has with respect to the medium. For the simplest arrangement comprising only two paths of equal length \(l\) the mean of speed of light in vacuum \(c_0\) is given by Eq.~(\ref{C.ZERO.TG}).
\begin{equation}
	c_{0}=\frac{2l}{t_{g}}
\label{C.ZERO.TG}
\end{equation}

The total propagation time there and back \(t_{g}\) can be exactly measured by a single clock.

\subsection{Synchronization method using light signals\label{SSEC:SYNCMETHLIGHT}}

The simplest arrangement for synchronization of two clocks in a system utilizes one single path in both directions. Consider a rod of lenght \(l\) with a light emitter at the end \(A\) and a mirror at the other end \(B\). On both ends are clocks with equal properties. The clock at \(A\) shows time \(t_{A}\), the clock at \(B\) shows time \(t_{B}\). A light signal starts at \(A\) at time \(t_{A(1)}\), travels to point \(B\), is reflected at \(B\) at time \(t_{B(2)}\), travels back and is absorbed at \(A\) at time \(t_{A(3)}\). The total propagation time \(t_{g}\) of the light pulse from emission at \(A\) to the absorption back at \(A\) is the difference of \(t_{A(3)}\) and \(t_{A(1)}\) (\ref{DELTA.TG2}).
\begin{equation}
	t_{g}=t_{A(3)}-t_{A(1)}
\label{DELTA.TG2}
\end{equation}

The propagation time \(t_{h}\) of the signal between emission at \(A\) and reflection at \(B\) is given by the difference of \(t_{B(2)}\) and \(t_{A(1)}\) (\ref{DELTA.TH.DEF}).
\begin{equation}
	t_{h}=t_{B(2)}-t_{A(1)}
\label{DELTA.TH.DEF}
\end{equation}

Because the speed of light \(c_{h}\) in this direction is unknown the time \(t_{B(2)}\) is unknown as well, and it is important to keep in mind that there are any assumptions possible for the instant of reflection as long as \(t_{B(2)}\) is within the limits \(t_{A(1)}\) and \(t_{A(3)}\). The instant of reflection has to be after the instant of emission, and the instant of absorption has to be after the instant of reflection (\ref{COND.TB}) in order to conserve the topology of time.
\begin{equation}
	t_{A(1)} \leq t_{B(2)} \leq t_{A(1)}
\label{COND.TB}
\end{equation}

For all further considerations it is useful to define the synchronization parameter \(\varepsilon\) given by the ratio of the propagation time \(t_{h}\) to the total propagation time \(t_{g}\) (\ref{EPS.DEF}).
\begin{equation}
	\varepsilon = \frac{t_{h}}{t_{g}} = \frac{t_{h}}{t_{h}+t_{r}}
\label{EPS.DEF}
\end{equation}

Due to condition (\ref{COND.TB}) \(\varepsilon\) must be within the limits 0 and 1. For more details see \cite[p.~51]{Mittelstaedt}.
\begin{equation}
	0 \leq \varepsilon \leq 1
\label{COND.EPS}
\end{equation}

Now, the clock at \(B\) is synchronized with the clock at \(A\) by choosing the time of reflection \(t_{B(2)}\) at \(B\) according to Eq.~(\ref{TB2}) with the synchronization parameter \(\varepsilon\) set to a particular value between 0 and 1. As long as \(\varepsilon\) meets condition (\ref{COND.EPS}) the results of synchronization comply with observable phenomena.
\begin{equation}
	t_{B(2)}=t_{A(1)} + \varepsilon \left( t_{A(3)}-t_{A(1)} \right)
\label{TB2}
\end{equation}

The synchronization parameter \(\varepsilon\) can also be expressed in terms of the speed of light. If we introduce the value \(c_{h}\) for the speed of light in the direction from \(A\) to \(B\), and the value \(c_{r}\) for the speed of light in the opposite direction according to Eqs.~(\ref{C.H}) and (\ref{C.R})
\begin{equation}
	c_{h}=\frac{l}{t_{h}}=\frac{l}{t_{B(2)}-t_{A(1)}}
\label{C.H}
\end{equation}

\begin{equation}
	c_{r}=\frac{l}{t_{r}}=\frac{l}{t_{A(3)}-t_{B(2)}}
\label{C.R}
\end{equation}

then from Eq.~(\ref{EPS.DEF}) the synchronization parameter \(\varepsilon\) is given by Eq.~(\ref{EPS.DEF.C}) featuring only the speed of light in the direction there and back of a given distance.
\begin{equation}
	\varepsilon=\frac{c_{r}}{c_{h}+c_{r}}
\label{EPS.DEF.C}
\end{equation}

The synchronization parameter \(\varepsilon\) is a measure for the anisotropy of speed of light in the direction given by the axis of the rod.
At the beginning of this section we talked about missing rules for the ratio \(v_h/v_r\). When substituting \(v_h\) by \(c_h\) and \(v_r\) by \(c_r\), the ratio can be expressed in terms of the synchronization parameter \(\varepsilon\). From Eq. (\ref{EPS.DEF.C}) we get the ratio \(c_h/c_r\) for the speed of light in the direction there and back, Eq. (\ref{RATIO:CH:CR}).
\begin{equation}
	\frac{c_h}{c_r} = \frac{1 - \varepsilon}{\varepsilon}
\label{RATIO:CH:CR} 
\end{equation}

Thus, getting a definition for the synchronization parameter \(\varepsilon\) is equivalent in getting a rule for the ratio \(v_h/v_r\).

\subsection{Synchronization schemes}

We assume there is more than one inertial system, and all systems are moving at different velocities with respect to each other. Because there are no restrictions in chosing a value for the synchronization parameter \(\varepsilon\), except that given by cond.\ (\ref{COND.EPS}), any of these inertial systems can be declared to be the stationary system \(K_0\). To achieve isotropic speed of light in this particular system we have to synchronize the clocks according to Einstein synchronization with the synchronization parameter \(\varepsilon_E\) in every direction of space. For all other inertial frames \(K'\) we have generally the same freedom to choose a value for the synchronization parameter. Thus we have an infinite number of combinations for the synchronization parameters. We can randomly select a particular combination, or we can introduce a further condition or define a property of space, which determines a particular combination. We will denote a particular combination of synchronization parameters as a \emph{synchronization scheme}. A synchronization scheme yields a particular selection of synchronization parameters applicable to all involved reference systems. From all possible combinations or schemes, there are two of special interest:
\begin{enumerate}
	\item The first scheme, proposed by Albert Einstein, applies Einstein synchronization on every inertial system.
	\item The second scheme employs medium synchronization with a synchronization parameter \(\varepsilon (v, \chi)\) depending on velocity \(v\) of the moving inertial system with respect to the stationary system as well as on angle \(\chi\) of the direction of synchronization with respect to the velocity vector \(\bf v\) (see Fig.~\ref{fig:d0256}).
\end{enumerate}

The two schemes differ with respect to a medium. Theory of special relativity ignores a medium and deals without one, because general isotropy of speed of light doesn't lead to reasonable properties for a realistic medium. Medium synchronization accepts explicitly a realistic medium by adaptation the synchronization parameter \(\varepsilon\).

It may seem that medium synchronization violates the principle of relativity, but there is always given the possibility to select any particular inertial system and make it to the system at rest. Thus the principle of relativity is satisfied. There is no need to declare all inertial systems to systems at rest at the same time, when obviously moving systems are observable. We can do this, and we are allowed to do this, but in this case we shouldn't wonder about paradox results, which are actually no paradoxa, they are real consequencies of what we do:---An observer in an Einstein synchronized system comparing his measuring-rod with measuring-rods of other inertial systems, which are moving with respect to him will always find his measuring-rod to be the longest. At the same time another observer in a moving system with respect to the first system but also Einstein synchronized will find his measuring-rod to be the longest of all. This is no contradiction and no paradox but an experimental measurable fact, which is only due to the synchronization scheme of clocks leading to the relativ simultaneity of events such as the events of measuring the end-positions of a moving rod.---A clock of a first inertial system meets a clock of a second inertial system only once, so there is no possibility given to find out, which one of the two clocks is running faster. Using light signals to achieve a second event for comparing the time readings of the two clocks yields a result depending on the synchronization scheme applied to the systems. When both systems are Einstein synchronized everyone of the two clocks can claim to run faster than the other. This seems to be paradox but it is actually measurable and therefore no contradiction. We do not need to endeavour general relativity to solve the paradoxa of special relativity, the results have to be taken as they are.

\subsubsection{Einstein synchronization}

When we assume isotropy of speed of light, i.e., the same magnitude for speed of light in both directions (from \(A\) to \(B\) as well as from \(B\) to \(A\)) we have to apply a synchronization parameter \(\varepsilon\) with the special value of \(0.5\). This method of synchronization is called Einstein synchronization, and we denote this specific synchronization parameter with subscript \(E\), Eq.~(\ref{EIN_SYNCH}).
\begin{equation}
	\varepsilon_E= 0.5
\label{EIN_SYNCH}
\end{equation}

When we speak of an Einstein synchronized inertial system we have applied condition (\ref{EIN_SYNCH}) not only on one particular direction but on every direction where clocks are to be synchronized in that system.

\subsubsection{Medium synchronization}

With medium synchronization we assign properties to space similar to those of a realistic medium. That is, isotropy of speed of light we find only in the system at rest. In all other moving inertial systems we have anisotropy according to the moving state with respect to the system at rest. The following example shall expound on that. -- A metal-rod with homogenious structure has isotropic speed of sound only in a frame of reference, where the rod is at rest. When using a frame of reference drifting along the axis of the rod at velocity \(v\) the speed of sound in the rod will be no more isotropic, but the rod will be still homogenious, nothing changed the property of the rod. That means, when we assume a homogenious structure for a given body or for space, isotropy of all properties cannot be concluded for all inertial systems of reference. This holds also for speed of light and electromagnetic waves in empty space.

Our job is to find an instruction for synchronization the clocks in moving systems. For that purpose we use the anisotropy of speed of light in the moving system, which is already measurable in the Einstein synchronized stationary system. In the moving system we have to adjust the clocks in such a manner, that the same anisotropy is yielded when measuring the speed of light in the moving system. We are enforced to this assumption because of the reasonable fact, that the relation of the time elapsed for a light pulse traveling along a given distance to the time elapsed traveling in the opposite direction should be always the same, no matter from which inertial system the time differences have been measured. This result could be confirmed specifically when additional signals of infinite speed were available.

\begin{figure}
	\centering
		\includegraphics{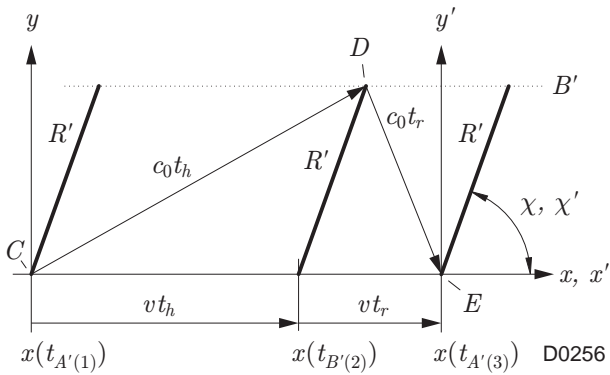}
	\caption{Medium synchronization, determination of synchronization parameter \(\varepsilon(v, \chi)\). Bold lines \(R'\) represent a rod at three instants \(t_{A'(1)}\), \(t_{B'(2)}\), and \(t_{A'(3)}\) measured with clocks of the stationary system \(K_0\). The rod represents the moving system \(K'\) and drifts at velocity \(v\) in the positive \(x\)-direction. Dotted line \(B'\) represents the path of end \(B'\) of the rod, the path of the other end \(A'\) is common to the \(x\)-axis. The position of the \(y'\)-axis of the moving system \(K'\) is depicted for time \(t_{A'(3)}\).---At time \(t_{A'(1)}\) a light pulse starts from end \(A'\) of the rod at position \(C\) and travels to the other end \(B'\) of the rod to position \(D\) (arrow \(c_0 t_h\)).---After elapsed time \(t_h\), at time \(t_{B'(2)}\), the light pulse is reflected by a mirror mounted at end \(B'\) of the rod and travels back to end \(A'\) (arrow \(c_0 t_r\)).---After further elapsed time \(t_r\), at time \(t_{A'(3)}\), the light pulse arrives at end \(A'\) of the rod at position \(E\).---Synchronization parameter \(\varepsilon(v, \chi)\) is calculated from propagation times \(t_h\) and \(t_r\) using Eq.~(\ref{EPS.DEF}).---Medium synchronization demands the value \(\varepsilon(v, \chi)\) to be applied in the moving system \(K'\), when two clocks \(T_{A'}\) and \(T_{B'}\), mounted anywhere on the moving rod \(R'\), are to be synchronized.---Please note that, due to lenght contraction, angle \(\chi'\) (measured in the moving system \(K'\)) is different from angle \(\chi\) (measured in the stationary system \(K_0\)).}
\label{fig:d0256}
\end{figure}

Consider a rod of length \(l\) with equal clocks on both ends \(A'\) and \(B'\) and the axis of the rod having an angle \(\chi\) with respect to the \(x\)-axis of the frame at rest (see Fig.~\ref{fig:d0256}). The rod is moving along the \(x\)-axis at speed \(v\). The magnitudes \(l\), \(\chi\), and \(v\) have been measured with clocks and measuring-rods of the stationary system \(K_{0}\). The end \(A'\) of the rod coincides with the origin of the coordinate system \(K'(x', y', z', t')\), where the rod is at rest.

At time \(t_{A'(1)}\) the origins of both coordinate systems \(K_{0}\) und \(K'\) are common. The \(x\)-, \(y\)- und \(z\)-axis of both coordinate systems are parallel to each other \((x \parallel x')\), \((y \parallel y')\), \((z \parallel z')\). A light ray emitted at time \(t_{A'(1)}\) at end \(A'\)  arrives at time \(t_{B'(2)}\) at end \(B'\) after elapsed time \(t_{h}\) according to Eq.~(\ref{DELTA.TH}).
\begin{equation}
	t_{h}=l \: \frac{v \cos{\chi} + \sqrt{c_{0}^2 - v^2 \sin^2{\chi}}}{c_{0}^2 - v^2}
\label{DELTA.TH}
\end{equation}

The light ray is reflected at time \(t_{B'(2)}\) at end \(B'\) and arrives back at end \(A'\) at time \(t_{A'(3)}\) after elapsed time \(t_{r}\) according to Eq.~(\ref{DELTA.TR}).
\begin{equation}
	t_{r}=l \: \frac{-v \cos{\chi} + \sqrt{c_{0}^2 - v^2 \sin^2{\chi}}}{c_{0}^2 - v^2}
\label{DELTA.TR}
\end{equation}

The synchronization parameter \(\varepsilon\) depends on velocity \(v\) and on angle \(\chi\), thus denoted by \(\varepsilon (v, \chi)\), and is calculated from (\ref{DELTA.TH}) and (\ref{DELTA.TR}) using Eq.~(\ref{EPS.DEF}).
\begin{equation}
	\varepsilon (v, \chi) = \frac{1}{2}
	\left ( 1 + \frac{v}{c_{0}}
	\frac {\cos{\chi}} {\sqrt{1 - \frac{v^2}{c_{0}^2}\sin^2{\chi}}}  \right)
\label{EPS.V.CHI}
\end{equation}

This is the equation for {\scshape medium synchronization}, Eq.~(\ref{EPS.V.CHI}), and gives the synchronization parameter \(\varepsilon\), applicable in a moving system \(K'\) drifting at velocity \(v\) in the positive \(x\)-direction, and the direction of the clock to be synchronized having an angle of \(\chi\) with the \(x\)-axis (Fig.~\ref{fig:d0256}). Both quantities, \(v\) and \(\chi\), have been measured in the system at rest \(K_0\).
Because it is much more convenient to measure angle \(\chi'\) in the moving system \(K'\), where the rod is at rest, we give the alternative Eq.~(\ref{EPS.V.CHI:PRIME}) for the synchronization parameter.
\begin{equation}
	\varepsilon (v, \chi') = \frac{1}{2}
	\left ( 1 + \frac{v}{c_0} \cos{\chi'} \right)
\label{EPS.V.CHI:PRIME}
\end{equation}

Equation (\ref{EPS.V.CHI:PRIME}) has been developed using the results of medium transformation tentatively, particularly the property of length contraction (see \ref{SSEC:LC}). When the drift velocity \(v\) is not available, the corresponding magnitude \(v'\) cannot be measured in the moving system, because the clocks aren't synchronized yet. But there is a way to overcome this problem. At first, the clocks in the moving system have to be Einstein synchronized. Then the drift velocity \(v'_E\) can be measured, this is the drift velocity of the system at rest with respect to the moving system, measured with Einstein synchronized clocks of the moving system. The drift velocity \(v\) is equally to the negative drift velocity \(v'_E\),  Eq.\ (\ref{MINUS:DRIFT}).
\begin{equation}
	v = -v'_E
	\label{MINUS:DRIFT}
\end{equation}

Subsequently, the clocks in the moving system can be medium synchronized according to Eq.\ (\ref{EPS.V.CHI:PRIME}). -- For an angle of \(\chi = 0\) Eq.~(\ref{EPS.V.CHI}) gets the simple form (\ref{EPS.V.ZERO}).
\begin{equation}
	\varepsilon \left(v, 0\right) = \frac{c_{0} + v } { 2 c_{0}}
	\label{EPS.V.ZERO}
\end{equation}

We use Eq. (\ref{EPS.V.ZERO}) when we consider the propagation of a light ray in the \(x\)-direction in order to derive the differential equation giving the dependence of primed time \(t'\) from the \(x\)-position.
-- For an angle of \(\chi = \pi / 2\) the cosine of \(\chi\) equals zero and the synchronization parameter has the constant value of \mbox{\(\varepsilon = 0.5\)}, Eq.~(\ref{EPS.05}), independent of velocity \(v\).
\begin{equation}
	\varepsilon \left(v, \frac{\pi}{2}\right) = 0.5
	\label{EPS.05}
\end{equation}

Equation (\ref{EPS.05}) is used for considering the propagation of a light ray perpendicular to the \(x\)-direction, i.e., either in the \(y\)- or \(z\)-direction.
\mbox{-- The} synchronization parameter calculated either from Eq.~(\ref{EPS.V.CHI}) or from Eq.~(\ref{EPS.V.CHI:PRIME}) is always within the limits 0 and 1 for any value of angle \(\chi\) or \(\chi'\) and for drift velocity \(v\), as long as the drift velocity is less than speed of light \(c_{0}\). Therefore \emph{medium synchronization generally complies with observable phenomena}. Hence, medium synchronization can be used without any further restriction, besides the already known limitation of the drift velocity \(v\).

\subsection{Effect of synchronization on the `now'-position}

\begin{figure}
	\centering
		\includegraphics{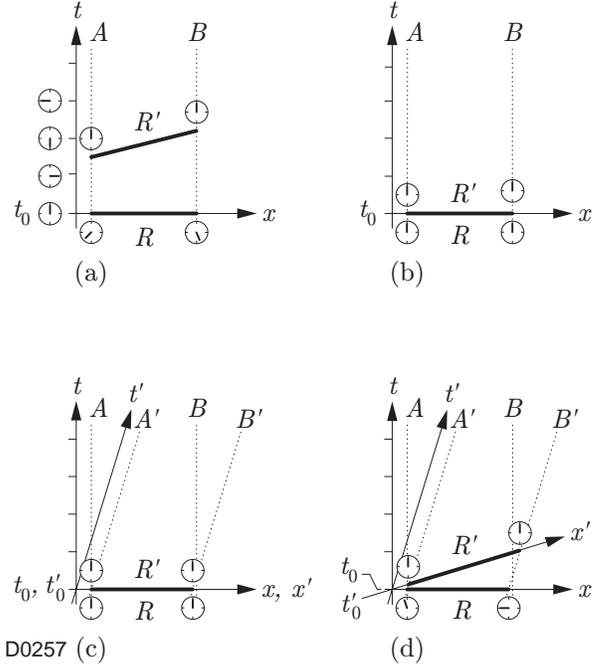}
	\caption{The effect of synchronization on the location shown by a rod in a space-time-like coordinate system with clocks on both ends. \(A\) and \(B\) denote world lines of the endpoints of the rod at rest, \(A'\) and \(B'\) denote world lines of the endpoints of the moving rod. Lines \(R\) indicate the positions of the rod at time \(t_{0}\) in the frame at rest \(K_{0}(x, t)\), clocks beyond the lines \(R\) show the rod-time at \(K_{0}\)-time \(t_{0}\). Lines \(R'\) indicate the position of the rod at rod-time \(t_{r0}\) represented by the clocks above the rod.---Fig.~\ref{fig:d0257}(a). Rod at rest with initially randomly distributed zero position settings of the clocks \(T_A\) and \(T_B\).---Fig.~\ref{fig:d0257}(b). The same rod with Einstein synchronized clocks \(T_A\) and \(T_B\).---Fig.~\ref{fig:d0257}(c). The rod from Fig.~\ref{fig:d0257}(b) moving at velocity \(v\) in the \(x\)-direction indicated by the sloped world lines \(A'\) and \(B'\). Clocks \(T_A\) and \(T_B\) are medium synchronized.---Fig.~\ref{fig:d0257}(d). The rod from Fig.~\ref{fig:d0257}(c) with Einstein synchronized clocks \(T_A\) and \(T_B\).---Length contraction and time dilation have been taken into account according to medium transformation and Lorentz transformation.}
\label{fig:d0257}
\end{figure}

What means synchronization, what happens at synchronization? Synchronization changes the zero position settings of the clocks in a system, so we have to look for a property, which is directly affected by the zero position settings in order to characterize the effect of synchronization. Such a property is the `now' of a system, the state of common time. Figure~\ref{fig:d0257} is showing a rod of arbitrary length equipped with two equally clocks \(T_A\) and \(T_B\) on both ends \(A\) and \(B\), respectively, thus representing its own system of reference \(K_{r}\left(x_{r}, t_{r}\right)\). The axis of the rod represents the \(x_{r}\)-axis of the system \(K_{r}\) and is aligned with the \(x\)-axis of our stationary system \(K_{0}\). We assume the rod beeing at rest in the system \(K_{0}\), and the clocks \(T_A\) and \(T_B\) having arbitrary zero position settings [Fig.~\ref{fig:d0257}(a)]. The position of the rod in the system \(K_{0}\) at \(K_{0}\)-time \(t_0\) is depicted by line \(R\), and the corresponding \(K_{r}\)-time readings are given by the clocks beyond the line. The line \(R'\) shows the `now'-position of the rod in the system \(K_{0}\) at \(K_{r}\)-time \(t_{r0}\).---By means of Einstein synchronization the clocks \(T_A\) and \(T_B\) with respect to the clock at the origin of the system at rest \(K_{0}\) the rod becomes part of the system \(K_{0}\) indicated by the same `now'-position of lines \(R\) and \(R'\) [Fig.~\ref{fig:d0257}(b)].---On acceleration the rod keeps in fact simultaneity with system \(K_{0}\) but moves in the \(x\)-direction at velocity \(v\) [Fig.~\ref{fig:d0257}(c)].---Applying Einstein synchronization again to the clocks \(T_A\) and \(T_B\) with respect to the clock at the origin of the moving system \(K'\), the `now' of the rod in this new moving system of reference \(K'\) is different from the `now' of the system at rest \(K_{0}\) [Fig.~\ref{fig:d0257}(d)], thus affecting the length of the same rod at rest when compared with the moving rod in the system \(K'\).

\section{Medium transformation\label{SEC:MT}}

With medium synchronization we are now able to develope the equations for transformation of the coordinates from our stationary system \(K_{0}\) into the moving system \(K'\). Magnitudes \(x\), \(y\), \(z\), and \(t\) are the position and time coordinates of the frame at rest \(K_{0}\), whereas \(x'\), \(y'\), \(z'\), and \(t'\) are the position and time coordinates of the moving frame \(K'\). The axes are assumed to be parallel \(( x \parallel x')\),  \(( y \parallel y')\), \(( z \parallel z)\) and the origin \(O'\) of the moving frame drifts at constant velocity \(v\) in the \(x\)-direction measured in the frame at rest. We assume, in the stationary system \(K_{0}\) the speed of light has magnitude \(c_{0}\) in every direction. A rod \(R'\) with two clocks \(T_A\) and \(T_B\) mounted on both ends \(A'\) and \(B'\), respectively, is aligned with the \(x'\)-axis. Then, the angle \(\chi'\) as well as angle \(\chi\) are zero and the synchronization parameter \(\varepsilon\) has the value given by Eq.\ (\ref{EPS.V.ZERO}). We use Eq.~(\ref{TB2}) with primed symbols \(t'\), \(A'\), and \(B'\) (the moving rod represents the moving system \(K'\)), plug in the synchronization parameter from (\ref{EPS.V.ZERO}),
\[
	{t'}_{B'(2)}={t'}_{A'(1)} + \frac{c_{0} + v } { 2 c_{0}} \left( {t'}_{A'(3)}-{t'}_{A'(1)}\right)
\]

and obtain.
\begin{equation}
	2 c_{0}\left( {t'}_{B'(2)}-{t'}_{A'(1)}\right)=\left(	c_{0} + v \right)
	\left( {t'}_{A'(3)}-{t'}_{A'(1)} \right)
	\label{TBTA.TATA}
\end{equation}

It is useful to subtract and add the time coordinate \({t'}_{B'(1)}\) on the left-hand side within the parentheses in order to get pure differential quotients for derivation of the differential equation.
\begin{eqnarray}
	2 c_{0}\left( {t'}_{B'(2)}-{t'}_{B'(1)}+{t'}_{B'(1)}-{t'}_{A'(1)}\right)\nonumber \\
	=\left(	c_{0} + v \right)
	\left( {t'}_{A'(3)}-{t'}_{A'(1)} \right)
	\label{TBTA.TATA2}
\end{eqnarray}

When we add the arguments of the frame at rest to time \(t'\), Eqs.~(\ref{TP.BP2}--\ref{TP.AP3})
\begin{eqnarray}
	{t'}_{B'(2)}&=&{t'}\left(l,0,0,t+\frac{l}{c_0-v}\right) \label{TP.BP2}\\
	{t'}_{B'(1)}&=&{t'}\left(l,0,0,t\right) \label{TP.BP1}\\
	{t'}_{A'(1)}&=&{t'}\left(0,0,0,t\right) \label{TP.AP1}\\
	{t'}_{A'(3)}&=&{t'}\left(0,0,0,t+\frac{l}{c_0-v}+\frac{l}{c_0+v}\right) \label{TP.AP3}
\end{eqnarray}

and let the distance \(l=\overline{A'B'}\) (measured in the frame at rest) go to zero, we get the differential equation (\ref{DTP.DX}) expressing the dependence of time \(t'\) on the \(x\)-coordinate of the frame at rest.
\begin{equation}
  \frac{\partial t'}{\partial x}=0
  \label{DTP.DX}
\end{equation}

In a similar way, regarding the propagation of light signals along the \(y\)- and \(z\)-axis, we get the dependence of time \(t'\) on the \(y\)- and \(z\)-coordinate, Eqs.\ (\ref{DTP.DY}) and (\ref{DTP.DZ}).
\begin{eqnarray}
  \frac{\partial t'}{\partial y}&=&0  \label{DTP.DY}\\
  \frac{\partial t'}{\partial z}&=&0  \label{DTP.DZ}
\end{eqnarray}

Eqs.~(\ref{DTP.DX}--\ref{DTP.DZ}) say, the time-coordinate \(t'\) is independent from the \(x\)-, \(y\)-, and \(z\)-coordinate of the frame at rest. The further derivation of medium transformation is similar to that given in \cite{Einst05}, except the transformation of the \(x\)-coordinate has to be performed in both directions, the positive and the negative direction. Comparing both results yields the ratio (\ref{REL.CH.CR}) between speed of light \(c'_h\) in the positive \(x'\)-direction and speed of light \(c'_r\) in the negative \(x'\)-direction of the moving system \(K'\). 
\begin{equation}
	\frac{c'_h}{c'_r}=\frac{c_0-v}{c_0+v}
	\label{REL.CH.CR}
\end{equation}

The variables \(c'_h\) and \(c'_r\) in the expressions for the propagation of a light pulse are eliminated by using again Eq.~(\ref{C.ZERO.TG}), leading to separate expressions for the variables \(c'_h\) and \(c'_r\) in terms of the vacuum speed of light \(c_0\) and the drift velocity \(v\).
\begin{equation}
	c'_h=\frac{c_0}{1+v/c_0}
	\label{CH:PRIME}
\end{equation}

\begin{equation}
	c'_r=\frac{c_0}{1-v/c_0}
	\label{CR:PRIME}
\end{equation}

In the following the ratio \(v/c_0\) frequently occures, so we define the beta factor given by (\ref{BETA}).
\begin{equation}
	\beta = \frac{v}{c_0}
	\label{BETA}
\end{equation}

With Eqs.~(\ref{CH:PRIME}) and (\ref{CR:PRIME}) the transformation of the \(x\)-coordinate reads
\begin{equation}
	x' = a(v)\: \gamma \left( x - vt \right),
	\label{X:PRIME:AV}	
\end{equation}

with the gamma factor
\begin{equation}
  \gamma = \frac{1}{\sqrt{1-\frac{v^2}{c_{0}^2}}}.
  \label{GAMMA}
\end{equation}

Analogously, we find the transformation of the other coordinates, Eqs.~(\ref{A:T:PRIME}--\ref{A:Z:PRIME}).
\begin{eqnarray}
  t'&=& a(v)\: \frac{1}{\gamma}t
  \label{A:T:PRIME} \\
  y'&=& a(v)\: y
  \label{A:Y:PRIME} \\
  z'&=& a(v)\: z
  \label{A:Z:PRIME}
\end{eqnarray}

The remaining scaling factor \(a(v)\) is determined --i-- comparing the results of two transformations, one in the positive and the other in the negative \(x\)-direction leading to Eq.~(\ref{AVMV})
\begin{equation}
	a(v) = a(-v),
	\label{AVMV}
\end{equation}

and --ii-- using a combined there and back transformation with the clocks in the moving system \(K'\) being Einstein synchronized before executing the back transformation. This obviously identical transformation yields Eq.~(\ref{AMVAV:1}).
\begin{equation}
	a(-v)a(v) = 1
	\label{AMVAV:1}
\end{equation}

Combining Eqs.~(\ref{AVMV}) and (\ref{AMVAV:1}) yields the equation \(a(v)a(v)=1\). Because the scaling factor \(a(v)\) must be greater than zero, otherwise time \(t'\) in the moving system is running backwards, the only possible value for the scaling factor \(a(v)\) is unity, and we attain the equations for {\scshape medium transformation}, (\ref{T.PRIME}--\ref{Z.PRIME}). Subscript \(M\) indicates variables pertaining to a medium synchronized system \(K'_M\).
\begin{eqnarray}
  t'_M&=&\frac{1}{\gamma}t
  \label{T.PRIME} \\
  x'_M&=&\gamma \left(x-vt\right)
  \label{X.PRIME} \\
  y'_M&=&y
  \label{Y.PRIME} \\
  z'_M&=&z
  \label{Z.PRIME}
\end{eqnarray}

Using the matrix notation from Sec.~\ref{SSS:COOTRANS} and comparing the coefficients of Eqs.~(\ref{T.PRIME}--\ref{Z.PRIME}) with the elements \(m_{ik}\) of matrix \(M\) in Eq. (\ref{MATRIX:EQ:TXYZ}), the matrix equation for medium transformation becomes
\begin{equation}
	\left(\begin{array}{c} c_0 t'_M	\\{\bf r}'_M \end{array}\right)
	= M
	\left(\begin{array}{c} c_0 t		\\{\bf r} \end{array}\right),
	\label{MATRIX:EQ:MEDIUM}
\end{equation}

with the elements of \(M\) given by Eq. (\ref{MATRIX:MT}).
\begin{equation}
	M	= \left(\begin{array}{cccc} 1 / \gamma  & 0			 & 0 & 0 \\
														  -\beta \gamma 	& \gamma & 0 & 0 \\
														  			0	  	& 0			 & 1 & 0 \\
														  			0	  	& 0			 & 0 & 1
					\end{array}\right)
	\label{MATRIX:MT}
\end{equation}

In literature this transformation is called a kind of a general Galilean transformation \cite[p.~137]{Mittelstaedt}, because it conserves simultaneity. Medium transformation not only takes care of a medium; it shows Galilean properties as well as relativistic properties and can be regarded as a kind of an intermediate transformation, so I suggest to use the name `medium transformation' for Eqs.~(\ref{T.PRIME}--\ref{Z.PRIME}).

\subsection{Properties of medium transformation}

Medium transformation has the same properties as Lorentz transformation regarding lenght contraction and time dilation. However, other properties have remarkable differences, i.e., theorem of velocity addition and simultaneity.

\subsubsection{Length contraction\label{SSEC:LC}}
The surface of a sphere with radius \(R'\), at rest in the moving system \(K'\), is given by Eq. (\ref{SPHERE:MOV}).
\begin{equation}
	x'^2 + y'^2 + z'^2 = R'^2
	\label{SPHERE:MOV}
\end{equation}

Applying medium transformation, Eqs. (\ref{T.PRIME}--\ref{Z.PRIME}), the equation of the surface at time \(t'=t=0\) becomes
\begin{equation}
	\left(\gamma x\right)^2 + y^2 + z^2 = R'^2 .
	\label{SPHERE}
\end{equation}
 
Comparing both equations, (\ref{SPHERE:MOV}) and (\ref{SPHERE}), shows that the radius \(R_x\) of the sphere, measured in the \(x\)-direction, is shortened to an amount of
\begin{equation}
	R_x = R'\sqrt{1-\frac{v^2}{c_0^2}},
	\label{L:CONTR}
\end{equation}

whereas the dimensions in the \(y\)- and \(z\)-direction remain constant.
\begin{equation}
	R_y = R_z = R'
\end{equation}

This phenomenon is called \emph{length contraction}.

\subsubsection{Time delation}

From Eq. (\ref{T.PRIME}) we can conclude directly that a clock in the moving system \(K'\) is running slower than a clock in the system at rest \(K_0\) by a factor of \(\sqrt{1-v^2/{c_0^2}}\). This phenomenon is known as \emph{time dilation}.

\subsubsection{Simultaneity}

Eq.~(\ref{T.PRIME}) shows no dependence of time on any coordinate of space. Thus two events occuring simultaneously in the system at rest on different positions are the same simultaneous in all other frames of reference. This is a major difference to Lorentz transformation.

\subsubsection{Theorem of velocity addition}

Theorem of velocity addition is given by Eqs.~(\ref{UX}--\ref{UZ}).
\begin{eqnarray}
	u_{x}&=&v+\frac{1}{\gamma^2}u'_{x}
	\label{UX} \\
	u_{y}&=&\frac{1}{\gamma}u'_{y}
	\label{UY} \\
	u_{z}&=&\frac{1}{\gamma}u'_{z}
	\label{UZ}
\end{eqnarray}

where \(u'_{x}\), \(u'_{y}\), and \(u'_{z}\) are the velocity components measured in the moving frame and \(u_{x}\), \(u_{y}\), and \(u_{z}\) are the velocity components measured in the frame at rest. In the moving frame, there are obviously velocities possible greater than speed of light in empty space \(c_0\). But regarding a particular direction \({\bf r}_p\) any speed \(v({\bf r}_p)\) in this direction cannot be greater than speed of light \(c_p\left({\bf r}_p\right)\) in this direction.

\section{Lorentz transformation versus medium transformation\label{SEC:MT:LT}}

Lorentz transformation and medium transformation differ only in the transformation of time. Equation (\ref{T.PRIME.L}) 
\begin{equation}
  t'_L=\gamma \left( t - \frac{v}{c_{0}^2} x \right),
  \label{T.PRIME.L}
\end{equation}

together with formulae (\ref{GAMMA}) and (\ref{X.PRIME}--\ref{Z.PRIME}), represent the Lorentz transformation (\ref{MATRIX:EQ:LORENTZ})
\begin{equation}
	\left(\begin{array}{c} c_0 t'_L	\\{\bf r}'_L \end{array}\right)
	= L \left(\begin{array}{c} c_0 t \\{\bf r} \end{array}\right) ,
	\label{MATRIX:EQ:LORENTZ}
\end{equation}

with the matrix \(L\) given by
\begin{equation}
	L = \left(\begin{array}{cccc} \gamma	& -\beta \gamma	 & 0 & 0 \\
														  -\beta \gamma		& \gamma & 0 & 0 \\
														  			0	  	& 0			 & 1 & 0 \\
														  			0	  	& 0			 & 0 & 1
					\end{array}\right) .
	\label{MATRIX:LORENTZ}
\end{equation}

Subscript \(L\) indicates variabels pertaining to a Lorentz transformed system, i.e., an Einstein synchronized, moving system. We know exactly when to apply Lorentz transformation or medium transformation if we have two inertial systems \(I_1\) and \(I_2\) with coordinate systems \(K_1\) and \(K_2\), respectively. It depends on the synchronization scheme applied to the systems. However, what is the case, when an inertial system changes its moving state with respect to others by an acceleration interval? What kind of synchronization do we find after that acceleration interval? A simple experiment discussed below will give an answer.

\subsection{Experiment with moving system of clocks}

\begin{figure}
	\centering
		\includegraphics{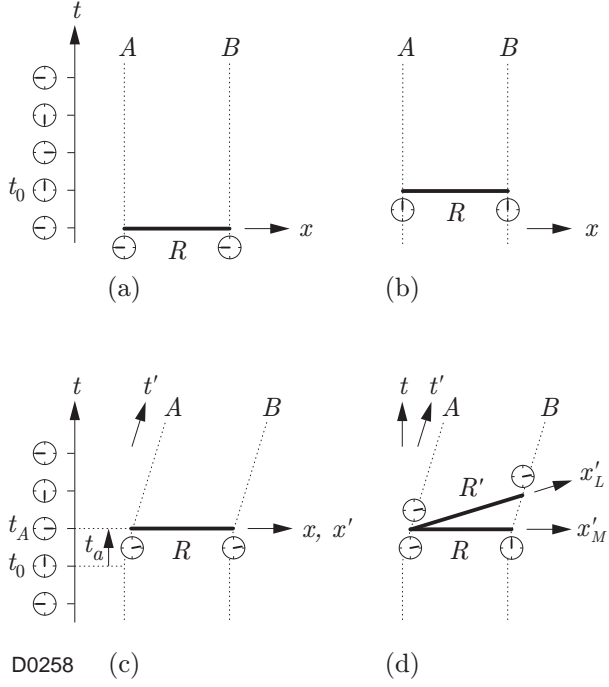}
	\caption{Experiment with moving system of clocks. Bold lines \(R\) represent the `now' of the rod at different instants in the stationary system \(K_0\).---Fig.~\ref{fig:d0258}(a). Rod at rest previous to acceleration with Einstein synchronized clocks \(T_A\) and \(T_B\) on both ends.---Fig.~\ref{fig:d0258}(b). Rod at the beginning of acceleration at \(K_0\)-time \(t_0\).---Fig.~\ref{fig:d0258}(c). Moving rod at \(K_0\)-time \(t_A\) after acceleration interval \(t_a\). The clocks \(T_A\) and \(T_B\) are in the state of medium synchronization.---Fig~\ref{fig:d0258}(d). Moving rod after acceleration interval \(t_a\) with clocks \(T_A\) and \(T_B\) again Einstein synchronized. Line \(R'\) indicates the `now' of the rod at \(K'\)-time \(t'_a\).}
\label{fig:d0258}
\end{figure}

Consider a rod of length \(l\) with two clocks \(T_A\) and \(T_B\) of equal properties mounted on both ends and the rod resting in the frame of reference (Fig.~\ref{fig:d0258}). The location of clock \(T_A\) is the origin of the coordinate system and the axis of the rod is aligned with the \(x\)-axis. The location of clock \(T_B\) has the coordinates \((l, 0, 0)\). The clocks are synchronized according to Einstein synchronization, thus behaving like a system at rest.

At time \(t_A = t_B = 0\) we accelerate the rod in the positive \(x\)-direction to velocity \(v\) and ask for the time the two clocks \(T_A\) and \(T_B\) should have to show immediately after the acceleration interval \(t_a\).

\subsection{Description with Lorentz transformation}

\begin{figure}
	\centering
		\includegraphics{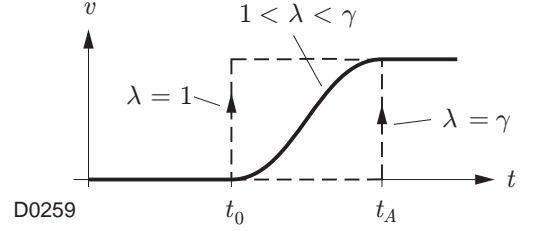}
	\caption{Expected range of correction factor \(\lambda\). When the total change in velocity \(v\) occures instantaneous at time \(t_0\), then \(\lambda\) will be unity. When the total change occures at time \(t_A\), then \(\lambda\) equals \(\gamma\). However, finite acceleration leads to a change in velocity \(v\) merely as depicted by the bold, solid line, so we expect the correction factor \(\lambda\) within the limits \mbox{1 and \(\gamma\).}}
\label{fig:d0259}
\end{figure}

Applying Eq.~(\ref{T.PRIME.L}) clock \(T_A\) has to show the time
\begin{equation}
	t'_A = \lambda \frac{1}{\gamma}t_a
	\label{TPR.A.L}
\end{equation}

and clock \(T_B\) has to show the time
\begin{equation}
	t'_B = \lambda \frac{1}{\gamma}t_a - \frac{v}{c_0^2}l
	\label{TPR.B.L}
\end{equation}

immediately after the acceleration interval \(t_a\). Thus clock \(T_B\) has a time lag of \(lv/c_0^2\) with respect to clock \(T_A\). The factor \(\lambda\) allows for compensation the nonconstant velocity \(v\) within the acceleration interval \(t_a\), because the \(\gamma\)-factor depends on velocity \(v\). The maximum value for \(\lambda\) is \(\gamma\), the minimum value is unity, cond.\ (\ref{LAMBDA.COND}) (see Fig.~\ref{fig:d0259}).
\begin{equation}
	1 \le \lambda \le \gamma
	\label{LAMBDA.COND}
\end{equation}

Condition (\ref{LAMBDA.COND}) doesn't take into account the effect of acceleration on the clocks due to the gravitation field the two clocks are exposed. According to general relativity a clock in a gravitation field is running slower than a clock in a space free of gravitation. Thus the correction factor \(\lambda\) may be less than unity but cannot be greater than \(\gamma\). Since we are looking only at the maximum value of \(\lambda\), the effect of acceleration on the clocks doesn't affect our considerations.

Even if we take the maximum value for \(\lambda\), for a given acceleration interval \(t_a\) setting Eq.~(\ref{TPR.B.L}) to zero we can always find a minimum length \(l_m\) so that after the acceleration interval clock \(T_B\) has to show a time of zero. If we make length \(l\) greater than \(l_m\) clock \(T_B\) has to run backwards, which is not reasonable. The other point is, clock \(T_B\) is running differently with respect to clock \(T_A\). Therefore we suppose the Lorentz transformation does not apply when describing the state of the moving system of clocks after the acceleration interval.

\subsection{Description with medium transformation}

With medium transformation the transformation of time has no space dependence. After the acceleration interval \(t_a\) clock \(T_A\) and clock \(T_B\) have to show the same time
\begin{equation}
	t'_A = t'_B = \lambda \frac{1}{\gamma}t_a.
	\label{TA.TB.M}
\end{equation}

The factor \(\lambda\) has the same meaning as in Eq.~(\ref{TPR.A.L}). There is no deviation between the time of clock \(T_A\) and the time of clock \(T_B\), which is much more reasonable than the result with Lorentz transformation. We find the moving system of clocks after the acceleration interval in a synchronization state according to medium transformation.

\subsection{Transformation according to Einstein synchronization}

What is to do to put the moving system of clocks into a state, which is correctly described by the Lorentz transformation? -- The two clocks \(T_A\) and \(T_B\) have to be Einstein synchronized once again by exchanging light signals. Is there a transformation given, which describes this Einstein synchronization correctly? -- When we suppose the moving system of clocks is in the state corresponding to medium transformation, then the way from the medium transformed system \(K'_M\) into the Lorentz transformed system \(K'_L\) is the following:

\begin{enumerate}
	\item Inverse medium transformation from the
moving coordinate system \(K'_M\) into the coordinate system at rest \(K_0\)
  \item Lorentz transformation of the coordinate system at rest \(K_0\) into the moving coordinate system \(K'_L\)
\end{enumerate}

The inverse medium transformation is given by Eqs.\ (\ref{T.TPM}--\ref{Z.ZPM}). Subscript `\(M\)' indicates variables of the medium transformed system to distinguish from variables of the Lorentz transformed system indicated by \mbox{subscript `\(L\)'}.
\begin{eqnarray}
	t&=&\gamma t'_M
	\label{T.TPM}	\\
	x&=&\frac{1}{\gamma}x'_M + \gamma v t'_M
	\label{X.XPMTPM} \\
	y&=&y'_M
	\label{Y.YPM}  \\
	z&=&z'_M
	\label{Z.ZPM}
\end{eqnarray}  
We plug in Eqs.~(\ref{T.TPM}--\ref{Z.ZPM}) into Lorentz transformation and obtain Eqs.~(\ref{ES.T}--\ref{ES.Z}), which we call Einstein synchronization as well.
\begin{eqnarray}
	t'_L&=&t'_M - \frac{v}{c_0^2}x'_M
	\label{ES.T} \\
	x'_L&=&x'_M
	\label{ES.X} \\
	y'_L&=&y'_M
	\label{ES.Y} \\
	z'_L&=&z'_M
	\label{ES.Z}
\end{eqnarray}

Einstein synchronization in matrix notation reads
\begin{equation}
	\left( 		\begin{array}{c} c_0 t'_L	\\{\bf r}'_L \end{array}\right)
	= E \left(\begin{array}{c} c_0 t'_M	\\{\bf r}'_M \end{array}\right)
	\label{MATRIX:EQ:EINSTEIN}
\end{equation}

with matrix \(E\) given by Eq.~(\ref{ES}).
\begin{equation}
	E
	= \left(\begin{array}{cccc} 1	& -\beta	& 0 & 0 \\
														  0	&		1 		& 0 & 0 \\
														  0	&		0			& 1 & 0 \\
														  0	&		0			& 0 & 1
					\end{array}\right)
	\label{ES}
\end{equation}

\subsection{Properties of Einstein synchronization}

\sloppy 

Einstein synchronization leaves the position coordinates constant, whereas the time coordinate gets the space dependence known from Lorentz transformation.

\subsection{Relation between Lorentz transformation, medium transformation and Einstein synchronization}
\fussy  

When we plug in Eq.~(\ref{MATRIX:EQ:MEDIUM}) into Eq.~(\ref{MATRIX:EQ:EINSTEIN}), we obtain an alternative equation for the Lorentz transformation featuring the matrices \(E\) and \(M\), Eq.~(\ref{MATRIX:EQ:LORENTZ:EM}).
\begin{equation}
	\left( 		\begin{array}{c} c_0 t'_L	\\{\bf r}'_L \end{array}\right)
	= E \: M \:
	\left(\begin{array}{c} c_0 t	\\{\bf r} \end{array}\right),
	\label{MATRIX:EQ:LORENTZ:EM}
\end{equation}

Comparing Eqs.~(\ref{MATRIX:EQ:LORENTZ:EM}) and (\ref{MATRIX:EQ:LORENTZ}) yields equation (\ref{LEM}) giving the matrix \(L\) for Lorentz transformation as the product of matrices \(E\) and \(M\)
\begin{equation}
	L = E \: M \: ,
	\label{LEM}
\end{equation}

and indeed, reckoning the product of the two matrices \(E\) and \(M\) from Eq.~(\ref{LT:COMP}) exactly yields the matrix \(L\) for Lorentz transformation.
\begin{equation}
	L
	= \left(\begin{array}{cccc} 1	& -\beta	& 0 & 0 \\
														  0	&		1 		& 0 & 0 \\
														  0	&		0			& 1 & 0 \\
														  0	&		0			& 0 & 1
					\end{array}\right)
		\left(\begin{array}{cccc} 1 / \gamma  & 0			 & 0 & 0 \\
														  -\beta \gamma		& \gamma & 0 & 0 \\
														  			0	  	& 0			 & 1 & 0 \\
														  			0	  	& 0			 & 0 & 1
					\end{array}\right)
	\label{LT:COMP}
\end{equation}

\begin{figure}
	\centering
		\includegraphics{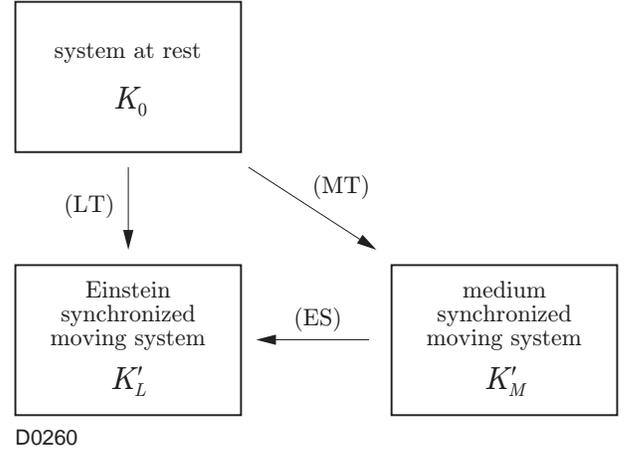}
	\caption{Relation between Lorentz transformation (LT), medium transformation (MT) and Einstein synchronization (ES). Lorentz transformation can be regarded as the product of medium transformation and Einstein synchronization.}
\label{fig:d0260}
\end{figure}

The matrix product has to be carried out in the usual way by taking rows from the left-hand matrix and columns from the right-hand matrix.
Equation (\ref{LEM}) says that Lorentz transformation can be taken as the product of two separate transformations, the medium transformation as the first part followed by Einstein synchronization (Fig.~\ref{fig:d0260}), which corresponds actually to observable phenomena, because when a system changes its state of movement we have always first to wait for the system to become inertial again and then we can apply Einstein synchronization on the clocks.

\section{Maxwell equations\label{SEC:MAXWELL}}

Now, we have to prove the properties of medium transformation regarding the compliance with observable phenomena by applying it on Maxwell equations. For convenience we use the homogenious version of the displacement current law (\ref{MAX.DC}), i.e., for empty space, and the induction law (\ref{MAX.IN}). First we apply medium transformation on both equations and then we derive the wave equation, because the properties of the tranformed Maxwell equations are demonstrated more easily by the wave equation. We have omitted the other two `divergence' laws, because they are not necessary for our purpose. Dots on the symbols are denoting time derivatives.
\def\MAXWELLDC
  {
		\rot {\bf B} = \frac{1}{c_0^2} \: {\dot {\bf E}}
  }
\begin{equation}
	\MAXWELLDC
\label{MAX.DC}
\end{equation}

\begin{equation}
	\rot {\bf E} = -\dot{\bf B}
\label{MAX.IN}
\end{equation}

\({\bf E}\) and \({\bf B}\) are the vectors of the electric and magnetic field in the system at rest. Applying medium transformation we obtain the extended Maxwell equations given by Eqs.~(\ref{MAX.DC.MT}) and (\ref{MAX.IN.MT}). The complete transformation of Eq.~(\ref{MAX.DC}) is given in appendix \ref{APP:MT:MAX:DC}.
\begin{equation}
	\rot {\bf B}' = \frac{1}{c_0^2} \frac{\partial}{\partial t'}
	 \left( {\bf E}' - {\bf v} \times {\bf B}' \right)
\label{MAX.DC.MT}
\end{equation}

\begin{equation}
	\rot {\bf E}' = - \frac{\partial}{\partial t'}
	 \left( {\bf B}'+ \frac{1}{c_0^2} {\bf v} \times {\bf E}' \right)
\label{MAX.IN.MT}
\end{equation}

\({\bf E}'\) and \({\bf B}'\) are the vectors of the electric and magnetic field in the moving system. There is no doubt, the medium transformation alters the form of the original Maxwell equations. But there is no need at all for the equations to be left constant by transformation as long as the transformed equations describe observables in a right manner. In order to show, that extended Maxwell equations (\ref{MAX.DC.MT}) and (\ref{MAX.IN.MT}) have similar properties than Maxwell equations for the frame at rest (\ref{MAX.DC}) and (\ref{MAX.IN}) we derive the wave equation for both cases.

\subsection{Wave equation and wave function for the frame at rest\label{SSEC:WAVE:REST}}

In this subsection we consider the properties of the electric field vector \(\bf E\) recommended from the wave equation for the frame at rest. This result we use in the following subsection for comparison with the properties of the field vector \({\bf E}'\) for the moving frame. Regarding the wave equations (\ref{WAVE.EQ}) and (\ref{WAVE.EQ.MT}) the wave functions for the magnetic field vectors \(\bf B\) and \({\bf B}'\) have identical properties. Hence, only the properties of the electric field are discussed, but the same statements apply for the magnetic field. -- From Eqs.~(\ref{MAX.DC}) and (\ref{MAX.IN}) we derive the wave equation (\ref{WAVE.EQ}) for the frame at rest
\begin{equation}
	c_0^2 \Delta {\bf E} - {\ddot{\bf E}} = c_0^2 \nabla \left( \nabla {\bf E} \right)
	\label{WAVE.EQ}
\end{equation}

On the left-hand side the vector of the electric field \({\bf E}\) occures seperately, whereas on the right-hand side it occures within the scalar product \((\nabla {\bf E})\). One possible solution of this differential equation is that both sides, the left-hand side as well as the right-hand side, have to be zero at the same time when we plug in a wave function. The wave function (\ref{WAVE.FUNC}) for a plane electromagnetic wave solves the left-hand part of Eq.~(\ref{WAVE.EQ}).
\begin{equation}
	{\bf E}({\bf r}, t)= {\bf E}_0 \sin \left(\omega_0 t - {\bf k} \SCDOT {\bf r} \right)
	\label{WAVE.FUNC}
\end{equation}

The amplitude vector \({\bf E}_0\) gives the amplitude \(E_0\) and the direction \(\widehat{\bf E}_0\) of the electric field and has components \(E_{0x}\), \(E_{0y}\), and \(E_{0z}\), Eq.~(\ref{E0}).
\begin{equation}
	{\bf E}_0 = \left( \begin{array}{c} E_{0y} \\ E_{0y} \\ E_{0z} \end{array} \right)
						= E_0 \, \widehat{\bf E}_0
	\label{E0}
\end{equation}

\(\omega_0\) is the angular frequency, \({\bf k}\) the wave vector, and \({\bf r}\) the vector from Eq.\ (\ref{POS.VEC.R}) to a point of space in the frame at rest. Wave vector \({\bf k}\), with wave number \(k\), has components \(k_x\),  \(k_y\), and \(k_z\), and a unit vector of direction \(\widehat{\bf k}\), all together given by Eq.\ (\ref{K:VECTOR}). 
\begin{equation}
			{\bf k} 
	     = \left(\begin{array}{c} k_x \\ k_y \\ k_z \end{array} \right)
	     =  k \,\left(\begin{array}{c}
	     							\widehat{k}_{x}\\ \widehat{k}_{y}\\ \widehat{k}_{z}
	     						\end{array} \right)
	     =  k \,\widehat{\bf k}
	\label{K:VECTOR}
\end{equation}

Wave number \(k\) is equally to the angular frequency \(\omega_0\) divided by speed of light \(c_0\), Eq. (\ref{K:OMEGA}).
\begin{equation}
	k = \frac{\omega_0}{c_0}
	\label{K:OMEGA}
\end{equation}

When setting the left-hand side of the wave Eq.~(\ref{WAVE.EQ}) to zero, generally the right-hand side is swept under the rug by the argument that the divergence of the electric field in vacuum is zero due to the absence of charges. But this argument does no more apply in our case, because we used almost the homogenious version of Maxwell equations implying the absence of charges. We cannot simply cancel the right-hand side of our wave equation, the wave function has to solve both, the left-hand side as well as the right-hand side of the wave equation. And in fact, the right-hand side gives information on the wave function to be found as will be shown below. The evidence for doing so becomes quite clear when we are dealing with the wave equation for the moving system.

As mentioned above, setting both sides of the wave Eq.~(\ref{WAVE.EQ}) to zero is one particular solution, other solutions are possible as well with wave functions making both sides of the wave equation equal but not zero, however, this kind of solutions will not be discussed here.
-- When we set the right-hand side of Eq.~(\ref{WAVE.EQ}) to zero, cancel the term \(c_0^2\nabla\), and plug in the wave function from Eq.~(\ref{WAVE.FUNC}), we obtain condition (\ref{NABLA_E_ZERO}).
\begin{equation}
	\nabla \cdot {\bf E} ({\bf r}, t)= 0
\label{NABLA_E_ZERO}
\end{equation}

Executing the \(\nabla\)-operator on the wave function \({\bf E}({\bf r},t)\) from Eq.~(\ref{WAVE.FUNC}) finally leads to condition (\ref{WAVE.VEC}).
\begin{equation}
	{\bf k} \cdot {\bf E}_0= 0 
	\label{WAVE.VEC}
\end{equation}

The propagation vector \({\bf c}_p\) of the electromagnetic wave from wave function (\ref{WAVE.FUNC}) is given by Eq. (\ref{CP.VEC}).
\begin{equation}
		{\bf c}_p
	=     \left(\begin{array}{c} c_{px} \\ c_{py} \\ c_{pz}	\end{array} \right)
	= c_0 \left(\begin{array}{c}
								\widehat{c}_{px} \\ \widehat{c}_{py} \\ \widehat{c}_{pz}
							\end{array} \right)
	= c_0 \widehat{{\bf c}}_p
	\label{CP.VEC}
\end{equation}

The coordinates \(\widehat{c}_{px}\), \(\widehat{c}_{py}\), and \(\widehat{c}_{pz}\) of the unit vector \(\widehat{{\bf c}}_p\) are also known as cosines of direction. 
In the frame at rest \(K_0\) wave vector \(\bf k\) has always the same direction as propagation vector \({\bf c}_p\), i.e., (\({\bf k} \parallel {\bf c}_p\)), see Fig.~\ref{fig:d0261}. In other words, the corresponding unit vectors \(\widehat{{\bf k}}\) and \(\widehat{{\bf c}}_p\) are equally, Eqs. (\ref{UVK:UVCP}) and (\ref{UVK:UVCP:COMP}).
\begin{eqnarray}
			\widehat{{\bf k}} & = & \widehat{{\bf c}}_p \label{UVK:UVCP} \\
	    \nonumber \\ 
	    \left(\begin{array}{c} \widehat{k}_{x} \\ \widehat{k}_{y} \\ \widehat{k}_{z}
							\end{array} \right)
							& = &
	    \left(\begin{array}{c} \widehat{c}_{px} \\ \widehat{c}_{py} \\ \widehat{c}_{pz}
							\end{array} \right) \label{UVK:UVCP:COMP}
\end{eqnarray}

Thus, condition (\ref{WAVE.VEC}) can be written as
\begin{equation}
	{\bf c}_p \cdot {\bf E}_0 = 0 \hspace{7 mm}\Leftrightarrow
	 \hspace{7 mm} {\bf E}_0\, \bot \,{\bf c}_p,
	\label{CP.E.ZERO}
\end{equation}

meaning that the field vector \({\bf E}_0\) has to be perpendicular to the propagation vector \({\bf c}_p\). The same holds for the field vector \({\bf B}_0\). This property of the wave equation (\ref{WAVE.EQ}) we will keep in mind for comparison with the corresponding property (\ref{CPP.EP.ZERO}) of the wave equation for the moving frame (\ref{WAVE.EQ.MT}).

\subsection{Wave equation and wave function for a moving frame\label{SSEC:WAVE:MOVING}}

From Eqs.~(\ref{MAX.DC.MT}) and (\ref{MAX.IN.MT}) we derive the extended wave Eq.~(\ref{WAVE.EQ.MT}) for a real moving frame (see appendix \ref{APP:DER:WAVE:EQ}).
\begin{eqnarray}
		c_0^2 \Delta {\bf E}'
	- \frac{1}{\gamma^2}{\ddot{\bf E}}'
	+ 2 \left({\bf v} \nabla \right){\dot{\bf E}}' = \hspace{20mm}\ \nonumber \\
		\frac{1}{c_0^2}{\bf v}\left( {\bf v} {\ddot{\bf E}}' \right) 
	+ \nabla \left( {\bf v} {\dot{\bf E}}' \right) 
	+ {\bf v} \left( \nabla {\dot{\bf E}}' \right)
 + c_0^2 \nabla \left( \nabla {\bf E}' \right)
\label{WAVE.EQ.MT} 
\end{eqnarray}

\begin{figure}
	\centering
		\includegraphics{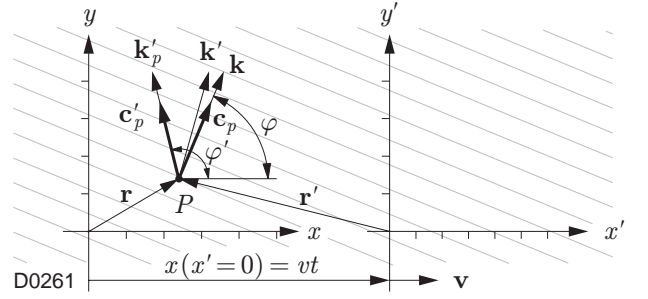}
	\caption{A plane electromagnetic wave is represented by gray lines indicating locations of common phase. Vectors \(\bf r\) and \({\bf r}'\) identify the \emph{same} point of space \(P\) where electromagnetic processes take place. For the frame at rest \(K_0\) wave vector \({\bf k}\) is parallel to the propagation vector \({\bf c}_p\). Both vectors are perpendicular to the wave front and have angle \(\varphi\) with the drift vector \(\bf v\). Wave vector \({\bf k}'\) and propagation vector \({\bf c}'_p\) of the moving frame \(K'\) have rather different directions, and none of them is perpendicular to the wave front. Propagation vector \({\bf c}'_p\) has angle \(\varphi'\) with drift vector \(\bf v\). The relation between angle \(\varphi'\) and angle \(\varphi\) is given by the aberration law (\ref{ABERRATION}). Tic marks on the axes indicate the length scale used. The length scale of the \(x'\)-axis is contracted, so the wave count per length unit in the moving system \(K'(x', y')\) is less than the wave count in the stationary system \(K_0(x, y)\), thus the wave vector \({\bf k}'\) is shifted slightly backwards towards the negative \(x\)-direction with respect to the wave vector \({\bf k}\), but is not affected by the drift vector \({\bf v}\) in opposite to the propagation vector \({\bf c}_p\).}
\label{fig:d0261}
\end{figure}

Eq.~(\ref{WAVE.EQ.MT}) makes clear that the right-hand side of the wave equation cannot be canceled by the argument with the divergence being zero. The first term on the right-hand side will never vanish and has to be cancelled by the other three terms when we plug in a wave function. Similar to Eq.~(\ref{WAVE.EQ}) there exist solutions when the left-hand part of  Eq.~(\ref{WAVE.EQ.MT})
\begin{equation}
	c_0^2 \Delta {\bf E}' - \frac{1}{\gamma^2}{\ddot{\bf E}}'
	                      + 2 \left({\bf v}\nabla \right){\dot{\bf E}}' = {\bf 0},
  \label{WAVE.EQ.MT.LEFT}	                      
\end{equation}

as well as the right-hand part of Eq.~(\ref{WAVE.EQ.MT})
\begin{equation}
	\frac{1}{c_0^2}{\bf v}\left( {\bf v}{\ddot{\bf E}}' \right) 
	+ \nabla \left( {\bf v}{\dot{\bf E}}' \right) 
	+ {\bf v} \left( \nabla {\dot{\bf E}}' \right)
 + c_0^2 \nabla \left( \nabla {\bf E}' \right) = {\bf 0}
 \label{WAVE.EQ.MT.RIGHT}
\end{equation}

are zero at the same time. Eq.~(\ref{WAVE.EQ.MT.LEFT}) is the wave equation for a moving frame, which is solved by the wave function (\ref{WAVE.FUNC.MT})
\begin{equation}
	{\bf E}'({\bf r}', t')= {\bf E}'_0 \sin \left( {\bf k}' \SCDOT \: {\bf c}'_p \ t'
	- {\bf k}' \SCDOT \: {\bf r}' \right)
	\label{WAVE.FUNC.MT}
\end{equation}

for a plane electromagnetic wave. \({\bf k}'\) is the wave vector, \({\bf c}'_p\) is the propagation vector and \({\bf r}'\) is the vector to a point of space in the moving frame. Eq.~(\ref{WAVE.FUNC.MT}) is obtained from Eq.~(\ref{WAVE.FUNC}) by applying medium transformation and is the general form of a function for a plane wave, holding good for a moving frame as well as for a frame at rest. Please note that in a real moving frame the wave vector \({\bf k}'\) and the propagation vector \({\bf c}'_p\) generally do not have the same direction, except when the wave propagates in the \({\bf v}\)-direction (see Fig.~\ref{fig:d0261}).

To achieve the propagation vector \({\bf c}'_p\) for the moving frame of reference, we have to apply theorem of velocity addition Eqs.~(\ref{UX}--\ref{UZ}) to the components of the propagation vector \({\bf c}_p\) from Eq.~(\ref{CP.VEC}). The propagation vector \({\bf c}'_p\) of the moving frame then becomes
\begin{equation}
	{\bf c}'_p 
	= c'_p \left(	\begin{array}{c}
								\widehat{c}'_{px} \cr \widehat{c}'_{py} \cr \widehat{c}'_{pz}
								\end{array} \right),
	\label{CP.VEC.MT}
\end{equation}

with a magnitude of
\begin{equation}
	c'_p = \gamma^2 c_0 \left(1 - \frac{v}{c_0}\widehat{c}_{px} \right),
	\label{CP.MAG.MT}
\end{equation}

and the coordinates of direction given by Eqs.~(\ref{CPE.X.MT}--\ref{CPE.Z.MT}).
\begin{eqnarray}
	\widehat{c}'_{px}&=& \frac{\widehat{c}_{px} - \frac{v}{c_0} }  
														{1 - \frac{v}{c_0}\widehat{c}_{px}}
	\label{CPE.X.MT} \\
	\widehat{c}'_{py}&=& \frac{\widehat{c}_{py}}
														{\gamma \left( 1 - \frac{v}{c_0}\widehat{c}_{px} \right) }
	\label{CPE.Y.MT} \\
	\widehat{c}'_{pz}&=& \frac{\widehat{c}_{pz}}
														{\gamma \left( 1 - \frac{v}{c_0}\widehat{c}_{px} \right) }
	\label{CPE.Z.MT}
\end{eqnarray}

The propagation vector \({\bf c}_p\) has angle \(\varphi\) with drift vector \({\bf v}\) in the system at rest, thus the \(x\)-coordinate of unit vector \(\widehat{\bf c}_p\) equals the cosine of angle \(\varphi\). With respect to Eq. (\ref{UVK:UVCP:COMP}) we get the relation (\ref{UV:CPX}) between cosines of direction for wave vectors \(\widehat{\bf k}\) and \(\widehat{\bf c}_p\) to angle \(\varphi\). 
\begin{equation}
  \widehat{k}_{x} = \widehat{c}_{px} = \cos \varphi
  \label{UV:CPX}
\end{equation}

Similarly, the propagation vector \({\bf c}'_p\) has angle \(\varphi'\) with drift vector \({\bf v}\) in the moving system, thus the \(x'\)-co\-or\-di\-nate of unit vector \(\widehat{\bf c}'_p\) equals the cosine of angle \(\varphi'\), Eq. (\ref{UV:CPX:PR}).
\begin{equation}
  \widehat{c}'_{px} = \cos \varphi'
  \label{UV:CPX:PR}
\end{equation}

With Eqs. (\ref{UV:CPX}) and (\ref{UV:CPX:PR}) Eq.~(\ref{CPE.X.MT}) becomes the well-known aberration law (\ref{ABERRATION}).
\begin{equation}
  \cos \varphi' = \frac{\cos \varphi - \frac{v}{c_0} } 
                       {1 - \frac{v}{c_0}\cos \varphi}\
  \label{ABERRATION}
\end{equation}

Wave vector \({\bf k}'\) in the moving frame is given by Eq.\ (\ref{WAVE.VEC.MT})
\begin{equation}
  {\bf k}'
  = \left(	\begin{array}{c}
  						{k_x / \gamma} \\ k_y \\ k_z
  					\end{array}	\right)
	= k' \left(	\begin{array}{c}
							\widehat{k}'_{x} \\ \widehat{k}'_{y} \\ \widehat{k}'_{z}
	  					\end{array}	\right)	
	= k' \widehat{{\bf k}}' ,
	 \label{WAVE.VEC.MT}
\end{equation}

with a wave number \(k'\) of
\begin{equation}
	k' = k \sqrt{1 + \frac{v^2}{c_0^2}\widehat{k}_x^2}  
	\label{WAVE.MAG.MT}
\end{equation}

and a unit vector of direction \(\widehat{\bf k}'\) given by Eq.~(\ref{WAVE.UNIT.VEC}).
\begin{equation}
  \widehat{\bf k}' =
   \frac{1}{\sqrt{1 + \frac{v^2}{c_0^2}\widehat{k}_x^2}}  
	 \left(	\begin{array}{c}
	 					{\widehat{k}_{x}/ \gamma} \\ \widehat{k}_{y} \\ \widehat{k}_{z}
 					\end{array}	\right)	
	=
		\left(	\begin{array}{c}
							\widehat{k}'_{x} \cr \widehat{k}'_{y} \cr \widehat{k}'_{z}
  					\end{array}	\right)	
   \label{WAVE.UNIT.VEC}
\end{equation}

Now, we'll focus on the right-hand part of Eq.~(\ref{WAVE.EQ.MT}). According to Eq.~(\ref{NABLA_E_ZERO}) we assume the right-hand part gives information on the direction of the field vectors \({\bf E}'\) and \({\bf B}'\). Cancelling the term (\ref{CANCEL.TERM})
\begin{equation}
	\frac{1}{c_0} \frac{\partial}{\partial t'} {\bf v} + c_0 \nabla
	\label{CANCEL.TERM}
\end{equation}

in Eq.~(\ref{WAVE.EQ.MT.RIGHT}) yields Eq.~(\ref{WAVE.EQ.MT.RIGHT.RED})
\begin{equation}
	\frac{1}{c_0}\left( {\bf v}{\dot{\bf E}}' \right) 
	 + c_0 \left( \nabla {\bf E}' \right) = 0,
	 \label{WAVE.EQ.MT.RIGHT.RED}
\end{equation}

and plugging in the wave function (\ref{WAVE.FUNC.MT}), we get the condition (\ref{KPP.EP.ZERO})
\begin{equation}
	{\bf k}'_p \cdot {\bf E}'_0 = 0 ,
	\label{KPP.EP.ZERO}
\end{equation}

with propagation wave vector \({\bf k}'_p\) given by Eq.\ (\ref{KPP}).
\begin{equation}
	{\bf k}'_p = {\bf k}' - \frac{{\bf k}' \cdot {\bf c}'_p} {c_0^2} \ {\bf v}
	\label{KPP}
\end{equation}

The propagation wave vector \({\bf k}'_p\) is represented by wave number \(k'_p\) and unit vector of direction \(\widehat{\bf k}'_p\) according to Eq. (\ref{KPP:PRODUCT}) 
\begin{equation}
  {\bf k}'_p = k'_p \widehat{\bf k}'_p ,
   \label{KPP:PRODUCT}
\end{equation}

with wave number \(k'_p\) given by Eq. (\ref{KPP:NUMBER})
\begin{equation}
   k'_p = \gamma k \left(1 - \frac{v}{c_0}k_x \right)
   \label{KPP:NUMBER}
\end{equation}

and unit vector of direction \(\widehat{\bf k}'_p\) given by Eq.~(\ref{UV:KPP}).
\begin{equation}
  \widehat{\bf k}'_p = \frac{1}{\gamma \left(1 - \frac{v}{c_0}k_x \right)}
  	 \left(	\begin{array}{c}
	 					 \gamma \left( \widehat{k}_{x} - \frac{v}{c_0}\right) \\
	 					 \widehat{k}_{y} \\ \widehat{k}_{z}
 					\end{array}	\right)	
   \label{UV:KPP}
\end{equation}

Comparing Eq.~(\ref{UV:KPP}) with Eqs.~(\ref{CPE.X.MT}--\ref{CPE.Z.MT}) and with regard to Eq.~(\ref{UVK:UVCP:COMP}), we find the wave vector \({\bf k}'_p\) having the same direction as propagation vector \({\bf c}'_p\) \footnote{
It was actually the transition from Eq. (\ref{KPP.EP.ZERO}) to Eq. (\ref{CPP.EP.ZERO}), wich convinced me doing something 	reasonable. Up to here I had serious doubts whether the medium transformation yields equations having any valuable physical meaning.
\mbox{-- First} of all, medium transformation alters the form of Maxwell equations, a monstrosity of its own. Moreover, the transformed field vectors came out ambiguously. It was really a challenge to find out a reasonable form of the transformed quantities, regardless to the physical meaning. I'm rather sure that only using the old-fashioned notation and transformation in component form enabled me to recognize the additional cross product term in the extended Maxwell equations. The reason for doing so was that I tried to develope the medium transformation as close as possible to the method used by Albert Einstein in his paper from 1905 \cite{Einst05}.
\mbox{-- The} second problem was the question: what is the meaning of the extended Maxwell equations? It was clear, that these equations of its own say nothing, regardless of holding good or not. And I knew I'll have to show at least one reasonable application of the extended Maxwell equations, otherwise the medium transformation would have no chance to become accepted. A well-known and frequently used application is the derivation of the wave equation. So I decided to demonstrate the properties of the extended Maxwell equations by the wave equation. This is no problem with the original Maxwell equations for a frame at rest, but it was rather difficult to get the meaning of the terms of the \emph{moving} wave equation and to get the terms into a reasonable order.
\mbox{-- Once} having arranged the terms in the right manner, the applicable wave function had to be found, which was merely a minor problem. When applying Lorentz transformation on the wave function the aberration law is obtained (see \cite{Einst05}). But with medium transformation no aberration law occures, so I had to look for the reason and found that theorem of velocity addition applied on the propagation vector \({\bf c}_p\) yields the aberration law, which is actually, in retrospect, not surprisingly.
\mbox{-- At} least, the wave function had to solve the wave equation, and it did so, this wasn't the problem, but what was the meaning of the propagation wave vector \({\bf k}'_p\)? Trying and playing and arranging the components of this vector showed---all of a sudden---that the wave vector \({\bf k}'_p\) is parallel to the propagation vector \({\bf c}'_p\), meaning that the transformed vectors \({\bf E}'\) and \({\bf B}'\) have to be perpendicular to the direction of propagation, a property, which is observable in every system of reference, regardless if moving or at rest. Finally, with this knowledge, the doubts on being right or wrong more and more vanished.
}, thus condition (\ref{KPP.EP.ZERO}) can be written as
\begin{equation}
	{\bf c}'_p \cdot {\bf E}'_0 = 0 \hspace{7 mm}\Leftrightarrow
	 \hspace{7 mm} {\bf E}'_0\, \bot \,{\bf c}'_p .
	\label{CPP.EP.ZERO}
\end{equation}

Condition (\ref{CPP.EP.ZERO}) says that in the moving system the vector \({\bf E}'_0\) has to be perpendicular to the propagation vector \({\bf c}'_p\) in order to be part of a function for a plane electromagnetic wave, which is fully in accordance with observable phenomena as well as with Eq.~(\ref{CP.E.ZERO}) for the frame at rest.
-- Independent from condition (\ref{CPP.EP.ZERO}) it can be shown, that medium transformation applied on the amplitude vector \({\bf E}_0\) from Eq.~(\ref{E0}) yields the medium transformed amplitude vector \({\bf E}'_0\), which is, indeed, perpendicular to the propagation vector \({\bf c}'_p\), provided vector \({\bf E}_0\) has been perpendicular to the propagation vector \({\bf c}_p\) in the frame at rest \(K_0\). -- The results from the previous two subsections concerning the electric field \(\bf E\) hold good as well for the magnetic field \(\bf B\).

\subsection{Application of Einstein synchronization}

We will demonstrate the effect of Einstein synchronization on a system complying with medium transformation by the example of the extended wave Eq.~(\ref{WAVE.EQ.MT.LEFT}). The drift vector \(\bf v\) is assumed to have only one component in the \(x\)-direction, Eq.~(\ref{DRIFT:VECTOR})
\begin{equation}
	{\bf v} = \left(	\begin{array}{c}
											v_x \\ v_y \\ v_z
				  					\end{array}	\right)	
	        = \left(	\begin{array}{c}
	        						v \\ 0 \\ 0
				  					\end{array}	\right),
\label{DRIFT:VECTOR}
\end{equation}

thus the wave equation reads
\begin{eqnarray}              
	{c_0^2}\left(   \frac{\partial^2}{\partial {x'_M}^2}
	       + \frac{\partial^2}{\partial {y'_M}^2}
	       + \frac{\partial^2}{\partial {z'_M}^2} \right) {\bf E}'\hspace{23mm}\ \nonumber \\
	- \frac{1}{\gamma^2} {\ddot{\bf E}}'
  + 2 v\ \frac{\partial}{\partial x'_M}\ {\dot{\bf E}}' = {\bf 0}.\hspace{7.7 mm}
  \label{WAVE.EQ.MT.LEFT.COMP}
\end{eqnarray}              

Dots on the symbols denote derivatives of the primed time \(t'\), double dots twofold derivatives of primed time. Applying Einstein synchronization (\ref{ES.T}--\ref{ES.Z}) on Eq.~(\ref{WAVE.EQ.MT.LEFT.COMP}) yields Eq.~(\ref{WAVE.EQ.LEFT.COMP})
\begin{equation}
	{c_0^2}\left(   \frac{\partial^2}{\partial {x'_L}^2}
	       + \frac{\partial^2}{\partial {y'_L}^2}
	       + \frac{\partial^2}{\partial {z'_L}^2} \right) {\bf E}'
	- {\ddot{\bf E}}' = {\bf 0} ,
  \label{WAVE.EQ.LEFT.COMP}
\end{equation}

which is identical to the form of the frame at rest, the left-hand side of Eq.~(\ref{WAVE.EQ}). The differential operators for Einstein synchronization are developed using the complete differential, similarly to that given in app.\ \ref{APP:MT:MAX:DC}. Analogously, Einstein synchronization turns Maxwell Eqs.~(\ref{MAX.DC.MT}) and (\ref{MAX.IN.MT}) for a moving frame of reference into the form (\ref{MAX.DC}) and (\ref{MAX.IN}) for a frame at rest.

\section{Conclusions\label{SEC:CONCLUSION}}

We have demonstrated that medium transformation can be successfully applied on Maxwell equations and on the functions of the electromagentic field. Also the results from medium transformation are consistent, the medium transformed wave functions solve the medium transformed wave equations. Furthermore, the results comply with observable phenomena, so we can say, there is indeed a way out of the dilemma, the possibility is given to consider a realistic medium. There is an alternative to the principle of constant speed of light.

Regarding the discrepancy exposed in appendix \ref{APP:MOV:CLKS}, the principle of relativity and the principle of constant speed of light should be discussed. It is not recommended to abandon both principles, but their scope should be considered and restricted so that electrodynamics in the framework of medium transformation can be established.

There have been a lot of experiments in the past intended to prove and to confirm special relativity, and in turn to reject aether theories. One of those experiments is the Wilson, but Puccini exactly obtained the same result using the electrodynamics of extended Maxwell equations \cite[p.\ 13\ Eq.\ (31)]{Puccini:physics/0407096v1}. Hence, the Wilson experiment is not suitable to confirm special relativity, and it is rather indicated to check all other experiments for concordance with medium transformation.

If we take a particular inertial system and consider the difference of coordinate systems between Einstein synchronization and medium synchronization, we find identical position coordinates, only the zero position settings of the clocks are differently. The structure of empty space cannot depend on the zero position settings of our clocks. Thus, when with medium transformation the structure of space can be given by the model of a realistic medium, the same structure should be compatible with Lorentz transformation and with theory of special relativity. That means, a medium, a preferred frame of reference, is not in contradiction with theory of special relativity, although principle of constant speed of light demands isotropy of speed of light in every inertial system, because we have to remember that this general isotropy is manipulated, is forced by the synchronization scheme and not a naturally given phenomenon.

The disadvantage of medium transformation is that in any moving system of reference we have always to know the moving state with respect to the system at rest \(K_0\), i.e., we have to know the velocity vector \(\bf v\). Another disadvantage is the transformed equations become more complex. However, the advantage of medium transformation is that a medium can be treated as an absolutely normal medium with properties similar to real matter. Another advantage is, there are no paradoxa at all. And the paradoxa of theory of special relativity are actually no paradoxa but observable facts only due to the synchronization scheme for the clocks.

Lorentz transformation as well as medium transformation, both yield the same phenomena regarding length contraction and time dilation, although their axioms are rather different. In other words, neither the principle of constant speed of light nor the concept of a medium can be responsible for relativistic phenomena, i.e., length contraction and time dilation. There must be a presupposition common to both transformations, and indeed, when looking more closer on the constituting conditions, one identifies Eq.~(\ref{C.ZERO.TG}) to be responsible for all the relativitstic behaviour.

Medium transformation is an alternative to Lorentz transformation, but not an exclusive. Both apply and both are valuable, they are not in contradiction to each other. Of course, Lorentz transformation will remain the more commonly used transformation, because it is a symmetrical transformation and more convenient to use. Medium transformation becomes important when the properties of the medium are to be considered.

\section{Outlook\label{SEC:OUTLOOK}}

Medium transformation yields the presupposition to consider a realistic medium, but this is only the very first step to a following theory of a medium, which is compared with this contribution an enormously big task. Before starting this project, we will pay attention on accompaning subjects in order to check for other presuppositions or to achieve useful information for a theory of a medium. For example, medium transformation doesn't take into account gravity. From this point of view it is on the same level as theory of special relativity. General relativity is a powerful instrument when dealing with gravity, but Riemannian spacetime is rigidly Lorentz invariant. Medium transformation violates Lorentz invariance, thus we have to look for a spacetime compatible with medium transformation, but leading to equivalent results compared with general relativity (like length contraction and time dilation of medium transformation compared with theory of special relativity). Secondly, we will test a simple model of photons explaining the properties of dispersion and the process of reflection. The model has to describe the forces arising from temporary dipole charges of the photons when penetrating electric fields with an inhomogenious gradient. The results from this model will give information on a proposed model of a particle \footnote{
There is a model of a particle comprised of two light quanta \cite{Weiss:Wellenmodell}. It is based on a preferred frame of reference and explains length contraction, time dilation, relativistic energy (mass) and momentum as well as the De Broglie wavelength associated with the momentum of the particle.} regarding the stability due to the charge of the particle. When all necessary conditions seem to be fulfilled, it will be reasonable to start the whole project, the development of medium theory.




\appendix   

\section{Transformation of displacement current law \label{APP:MT:MAX:DC}}

We assume, the drift vector \(\bf v\) has only one component in the \(x\)-direction with magnitude \(v\) according to Eq.~(\ref{DRIFT:VECTOR}). The complete differential (\ref{DIFF_TOTAL})
\begin{equation}
	\frac{\partial}{\partial t}
	= \frac {\partial t'}{\partial t} \frac{\partial}{\partial t'} 
	+ \frac {\partial x'}{\partial t} \frac{\partial}{\partial x'} 
	+ \frac {\partial y'}{\partial t} \frac{\partial}{\partial y'} 
	+ \frac {\partial z'}{\partial t} \frac{\partial}{\partial z'},
		\label{DIFF_TOTAL}
\end{equation}

applied on Eqs.~(\ref{T.PRIME}--\ref{Z.PRIME}) then yields the differential operator (\ref{DIFF_OP_T}) applicable for medium transformation of time derivatives. Analogously, we obtain the differential operators for transformation of space derivatives, Eqs.\ (\ref{DIFF_OP_X}--\ref{DIFF_OP_Z}).
\begin{eqnarray}
	\frac{\partial}{\partial t}&=&\frac{1}{\gamma}\ \frac{\partial}{\partial t'}
		  - \gamma v \ \frac{\partial}{\partial x'}
		\label{DIFF_OP_T} \\
	\frac{\partial}{\partial x}&=&\gamma \ \frac{\partial}{\partial x'}
		\label{DIFF_OP_X} \\
	\frac{\partial}{\partial y}&=&\frac{\partial}{\partial y'}
		\label{DIFF_OP_Y} \\
	\frac{\partial}{\partial z}&=&\frac{\partial}{\partial z'}
		\label{DIFF_OP_Z}
\end{eqnarray}

Maxwell equation of displacement current for empty space, Eq.~(\ref{MAX.DC})
\def\theequation{\arabic{equation}}
\begin{equation}
  \MAXWELLDC, 
\tag{\ref{MAX.DC}}
\end{equation}
\def\theequation{\Alph{section}\arabic{equation}}

has the components
\begin{eqnarray}
	\frac{\partial B_z}{\partial y} - \frac{\partial B_y}{\partial z}&=&
  \frac{ 1 }{ c_0^2}\ \frac{\partial E_x }{ \partial t},
\label{MAX.DC.X}\\
	\frac{\partial B_x } { \partial z} - \frac{\partial B_z } { \partial x}&=&
  \frac{ 1 }{c_0^2}\ \frac{\partial E_y}{\partial t},  \hspace*{18mm}
\label{MAX.DC.Y}\\ {\rm and} \hspace{10mm}
	\frac{\partial B_y}{\partial x} - \frac{\partial B_x}{\partial y}&=&
  \frac{1}{c_0^2}\ \frac{\partial E_z}{\partial t}\: .
\label{MAX.DC.Z}
\end{eqnarray}

We start with Eq.~(\ref{MAX.DC.X}) and plug in the differential operators according to Eqs.~(\ref{DIFF_OP_T}--\ref{DIFF_OP_Z}).
\[ 
	\frac{\partial B_z } { \partial y'} - \frac{\partial B_y } { \partial z'} =
  \frac{ 1 } { c_0^2}\ \left( \frac{1 } { \gamma}\ \frac{\partial } { \partial t'}
	                           - \gamma v \ \frac{\partial } { \partial x'}
	                    \right) E_x
\] 

On the left-hand side we add the terms \(\partial E_y /{\partial y'}\) and \(\partial E_z /{\partial z'}\), each multiplied by the ratio \(v/c_0^2\), and subtract them again from the differential quotients of the magnetic field components with the corresponding differentials for \(y'\) and \(z'\), so that the changes cancel itself.
\begin{eqnarray}
    \frac{v } { c_0^2} \frac{\partial E_y } { \partial y'}
  + \frac{v } { c_0^2} \frac{\partial E_z } { \partial z'} \hspace{45mm}\ \nonumber \\
	+ \frac{\partial B_z } { \partial y'}
  - \frac{v } { c_0^2} \frac{\partial E_y } { \partial y'}
	- \frac{\partial B_y } { \partial z'}
	- \frac{v } { c_0^2} \frac{\partial E_z } { \partial z'} \hspace{10mm}\ \nonumber \\
  = \frac{1 } { c_0^2}\ \frac{1 } { \gamma}\ \frac{\partial E_x } { \partial t'}
  - \gamma \ \frac{v } { c_0^2} \frac{\partial E_x } { \partial x'} \nonumber
\end{eqnarray}

We transform the differential operators of the first two terms back into the coordinate system at rest \(K_0\), add the term \(\partial E_x /{\partial x}\), also multiplied by the ratio \(v/c_0^2\), and subtract it again at the same place in order to cancel the changes. We extract the ratio \(v/c_0^2\) from the three positiv differential quotients.
\begin{eqnarray}
  - \frac{v } { c_0^2} \frac{\partial E_x } { \partial x}
  + \frac{v } { c_0^2} \left(   \frac{\partial E_x } { \partial x}
                           	 + \frac{\partial E_y } { \partial y}
                             + \frac{\partial E_z } { \partial z}
                      \right)
    \hspace{15mm}\ \nonumber \\
	+ \frac{ {\partial \left( B_z -\frac{v } { c_0^2} E_y \right)} } { \partial y'}
	- \frac{ {\partial \left( B_y +\frac{v } { c_0^2} E_z \right)} } { \partial z'}
		\hspace{0mm}\ \nonumber \\
  = \frac{1 } { c_0^2}\ \frac{1 } { \gamma}\ \frac{\partial E_x } { \partial t'}
  - \gamma \ \frac{v } { c_0^2} \frac{\partial E_x } { \partial x'}
    \nonumber
\end{eqnarray}

The expression in the first parentheses is the `divergence' of the vector field and must be zero because of condition (\ref{NABLA_E_ZERO}). The first term on the left-hand side, transformed into the moving System \(K'\), equals the second term on the right-hand side of the equation, both terms cancel each other, thus we obtain Eq. (\ref{MAX_DC_XT}) for the \(x'\)-component of the first Maxwell equation.
\begin{eqnarray}
	  \frac{ {\partial \left( B_z -\frac{v } { c_0^2} E_y \right)} } { \partial y'}
	- \frac{ {\partial \left( B_y +\frac{v } { c_0^2} E_z \right)} } { \partial z'}
	  \hspace{15mm}\ \nonumber \\
  = \frac{1 } { c_0^2}\ \frac{1 } { \gamma}\ \frac {\partial E_x } { \partial t'}\hspace{10mm}\ 
\label{MAX_DC_XT}
\end{eqnarray}

Now, we take Eq.~(\ref{MAX.DC.Y}) and plug in the differential operators Eqs.~(\ref{DIFF_OP_T}--\ref{DIFF_OP_Z}).
\[
	   \frac{\partial B_x } { \partial z'}
	 - \gamma \frac{\partial B_z } { \partial x'}
	 = \frac{ 1 } { c_0^2}\ \left( \frac{1 } { \gamma}\ \frac{\partial } { \partial t'}
	                           - \gamma v \ \frac{\partial } { \partial x'}
	                       \right) E_y
\]

\[
	   \frac{\partial B_x } { \partial z'}
	 - \gamma \frac{\partial B_z } { \partial x'}
	 + \gamma \ \frac{v } { c_0^2}\frac{\partial E_y } { \partial x'}
	 = \frac{1 } { c_0^2 \gamma}\ \frac{\partial E_y } { \partial t'}
\]

\[
	   \frac{\partial B_x } { \partial z'}
	 - \frac{\partial \gamma \left( B_z - \frac{v } { c_0^2} E_y \right) } { \partial x'}
	 = \frac{1 } { c_0^2 \gamma^2}\ \frac{\partial \gamma E_y } { \partial t'}
\]

By resolution of the squared gamma factor with Eq.\ (\ref{GAMMA}) we obtain:
\begin{equation}
	   \frac{\partial B_x } { \partial z'}
	 - \frac{\partial \gamma \left(B_z - \frac{v } { c_0^2} E_y \right) } { \partial x'}
	 = \frac{1 } { c_0^2} \left(1-\frac{v^2 } { c_0^2} \right)
			\frac{\partial \gamma E_y } { \partial t'}
\label{MAX_DC_YPT}	 
\end{equation}

At this point, our project seems to be doomed to failure, because the `transformed' field component \(E_y\) on the right-hand side doesn't compare with the expected term \(\left( E_y - vB_z \right)\). We are looking for a solution by `enhancement' of the right-hand side. We split the term \(\left( 1 - v^2 / c_0^2 \right)\) and subtract and add the term
\[
  \frac{v } { c_0^2}\ \frac{\partial \gamma B_z } { \partial t'}
\]

on the right-hand side of Eq.~(\ref{MAX_DC_YPT}).
\begin{eqnarray}
	   \frac{\partial B_x } { \partial z'}
	 - \frac{\partial \gamma \left(B_z - \frac{v } { c_0^2} E_y \right) } { \partial x'}=
	 \hspace{30mm}\ \nonumber \\
	   \frac{1 } { c_0^2}\ \frac{\partial \gamma E_y } { \partial t'}
	 - \frac{v } { c_0^2}\ \frac{\partial \gamma B_z } { \partial t'}
	 + \frac{v } { c_0^2}\ \frac{\partial \gamma B_z } { \partial t'}
	 - \frac{1 } { c_0^2}\ \frac{v^2 } { c_0^2} \frac{\partial \gamma E_y } { \partial t'}
	 \nonumber
\end{eqnarray}

Now, the \(y\)-component of the transformed Maxwell equation reads as expected and an additional term with the magnetic component occures, leading to Eq.~(\ref{MAX_DC_YT}).
\begin{eqnarray}
	   \frac{\partial B_x } { \partial z'}
	 - \frac{\partial \gamma \left(B_z - \frac{v } { c_0^2} E_y \right) } { \partial x'}=
	 \hspace{30mm}\ \nonumber \\
	   \frac{1 } { c_0^2}\ \frac{\partial \gamma \left( E_y - vB_z \right) } { \partial t'}
	 + \frac{v } { c_0^2}\ 
	   \frac{\partial \gamma \left( B_z - \frac{v } { c_0^2} E_y \right) } { \partial t'}
		\hspace{5mm}\ \label{MAX_DC_YT}
\end{eqnarray}

Similar to the \(y\)-component we get the \(z\)-component from Eq.~(\ref{MAX.DC.Z}):
\begin{eqnarray}
	   \frac{\partial \gamma \left(B_y + \frac{v } { c_0^2} E_z \right) } { \partial x'}
	 - \frac{\partial B_x } { \partial y'}=
	 \hspace{30mm}\ \nonumber \\
	   \frac{1 } { c_0^2}\ \frac{\partial \gamma \left( E_z + vB_y \right) } { \partial t'}
	 - \frac{v } { c_0^2}\ 
	   \frac{\partial \gamma \left( B_y + \frac{v } { c_0^2} E_z \right) } { \partial t'}.
	 \hspace{5mm}\ \label{MAX_DC_ZT}
\end{eqnarray}

Using the equivalences
\begin{tabbing} \hspace{3.4mm}\= \hspace{40mm}\= \kill
\>\(E'_x = E_x\),\>
\(B'_x = B_x\),
\end{tabbing} \vspace{-3mm}
\begin{tabbing} \hspace{3.4mm}\= \hspace{40mm}\= \kill
\>\(E'_y = \gamma \left( E_y - vB_z \right)\),\> 
\(B'_y = \gamma \left( B_y + \frac {v } { c_0^2}E_z \right)\), 
\end{tabbing}
\begin{tabbing} \hspace{3.4mm}\= \hspace{40mm}\= \kill
\>\(E'_z = \gamma \left( E_z + vB_y \right)\), {\rm and}\>
\(B'_z = \gamma \left( B_z - \frac{v } { c_0^2} E_y \right)\),
\end{tabbing}

finally leads to Maxwell Eqs.~(\ref{MAX_DC_X'}--\ref{MAX_DC_Z'}) for a moving frame.
\begin{eqnarray}
	\frac{\partial B'_z } { \partial y'} - \frac{\partial B'_y } { \partial z'}&=&
  \frac{ 1 } { c_0^2}\ \frac{\partial E'_x } { \partial t'}
\label{MAX_DC_X'}\\
	\frac{\partial B'_x } { \partial z'} - \frac{\partial B'_z } { \partial x'}&=&
  \frac{ 1 } { c_0^2}\ \frac{\partial \left( E'_y + vB'_z \right) } { \partial t'}
\label{MAX_DC_Y'}\\
	\frac{\partial B'_y } { \partial x'} - \frac{\partial B'_x } { \partial y'}&=&
  \frac{ 1 } { c_0^2}\ \frac{\partial \left( E'_z - vB'_y \right) } { \partial t'}
\label{MAX_DC_Z'}
\end{eqnarray}

From the equations in component form (\ref{MAX_DC_X'}--\ref{MAX_DC_Z'}) we `guess' the Maxwell equation of displacement current for vacuum in vectorial form, Eq.~(\ref{MAX.DC.MT}).
\def\theequation{\arabic{equation}}
\begin{equation}
	\rot {\bf B}' = \frac{1}{c_0^2} \frac{\partial}{\partial t'}
	 \left( {\bf E}' - {\bf v} \times {\bf B}' \right)
	\tag{\ref{MAX.DC.MT}}
\end{equation}
\def\theequation{\Alph{section}\arabic{equation}}

Within Eq.~(\ref{MAX.DC.MT}), written in components (\ref{MAX_DC_X'L}--\ref{MAX_DC_Z'L}), we find additional terms compared with Eqs.~(\ref{MAX_DC_X'}--\ref{MAX_DC_Z'}). 
\begin{equation}
	\frac{\partial B'_z } { \partial y'} - \frac{\partial B'_y } { \partial z'} = 
  \frac{ 1 } { c_0^2}\ \frac{\partial } { \partial t' } 
    \left[E'_x - \left( v_y B'_z - v_z B'_y \right) \right]
\label{MAX_DC_X'L}
\end{equation}
\vspace{-5mm}
\begin{equation}
	\frac{\partial B'_x } { \partial z'} - \frac{\partial B'_z } { \partial x'} = 
  \frac{ 1 } { c_0^2}\ \frac{\partial } { \partial t' } 
    \left[E'_y - \left( v_z B'_x - v_x B'_z \right) \right]
\label{MAX_DC_Y'L}
\end{equation}
\vspace{-5mm}
\begin{equation}
	\frac{\partial B'_y } { \partial x'} - \frac{\partial B'_x } { \partial y'} = 
  \frac{ 1 } { c_0^2}\ \frac{\partial } { \partial t' } 
    \left[E'_z - \left( v_x B'_y - v_y B'_x \right) \right]
\label{MAX_DC_Z'L}
\end{equation}

Our vector of drift velocity \(\bf v\) used for the differential operators has only one component in the \(x\)-direction according to Eq.~(\ref{DRIFT:VECTOR}). When we plug in these components into Eqs.~(\ref{MAX_DC_X'L}--\ref{MAX_DC_Z'L}) we obtain exactly Eqs.~(\ref{MAX_DC_X'}--\ref{MAX_DC_Z'}). Thus our guess seems to be verified. -- A similar procedure applied on Eq.~(\ref{MAX.IN}) yields the medium transformed induction law (\ref{MAX.IN.MT}).

\section{Derivation of wave equation for the moving frame\label{APP:DER:WAVE:EQ}}

The extended Maxwell equations of displacement current (\ref{MAX.DC.MT}) and induction law (\ref{MAX.IN.MT}) for empty space read:
\def\theequation{\arabic{equation}}
\begin{equation}
	\rot {\bf B}' = \frac{1}{c_0^2} \frac{\partial}{\partial t'}
	\left( {\bf E}' - {\bf v} \times {\bf B}' \right)
	\tag{\ref{MAX.DC.MT}}
\end{equation}  
\begin{equation}
	\rot {\bf E}' = - \frac{\partial}{\partial t'}
	\left( {\bf B}'+ \frac{1}{c_0^2} {\bf v} \times {\bf E}' \right)
	\tag{\ref{MAX.IN.MT}}
\end{equation}
\def\theequation{\Alph{section}\arabic{equation}}

From these two equations we derive the wave equation for the electic field vector \(\bf E'\). The rotor operator applied on Eq.~(\ref{MAX.IN.MT}) yields
\[
	   \rot \rot {\bf E}' =
	 - \frac{\partial } { \partial t'}\rot {\bf B}'
	 - \frac{1 } { c_0^2} \: \frac{\partial } { \partial t'}\rot 
	   \left( {\bf v} \times {\bf E}' \right).
\]

Substituting \(\rot {\bf B}'\) by the right-hand side of Eq.~(\ref{MAX.DC.MT}) yields
\begin{eqnarray}
	   \rot \rot {\bf E}' =
	 - \frac{\partial } { \partial t'} 
	     \left[  \frac{1 } { c_0^2} \frac{\partial } { \partial t'}
	             \left( {\bf E}' - {\bf v} \times {\bf B}' \right)
	     \right]
	 \hspace{0mm}\ \nonumber \\
	 - \frac{1 } { c_0^2} \: \frac{\partial } { \partial t'}\rot 
	   \left( {\bf v} \times {\bf E}' \right) .
	 \hspace{10mm}\ \nonumber
\end{eqnarray}

We replace the symbol `rot' by `\(\nabla \times\)' when it is appropriate,
\begin{eqnarray}
	   \rot \rot {\bf E}' =
	 - \frac{\partial } { \partial t'} 
	     \left[  \frac{1 } { c_0^2} \frac{\partial } { \partial t'}
	             \left( {\bf E}' - {\bf v} \times {\bf B}' \right)
	     \right]
	 \hspace{0mm}\ \nonumber \\
	 - \frac{1 } { c_0^2} \: \frac{\partial } { \partial t'}\: \nabla \times \: 
	   \left( {\bf v} \times {\bf E}' \right),
	 \hspace{10mm}\ \nonumber
\end{eqnarray}

\vspace*{0mm}
and resolve twofold cross products by means of the identity (\ref{DCROSS}),
\begin{equation}
	{\bf a} \times \left( {\bf b} \times {\bf c} \right)
	= {\bf b} \left( {\bf ac} \right) 
	- {\bf c} \left( {\bf ab} \right) ,
\label{DCROSS}	
\end{equation}

whereas the vectors within scalar products commute \(({\bf ab})=({\bf ba})\). Dots on the symbols denote derivatives of the primed time \(t'\), double dots twofold derivatives of primed time.
\begin{eqnarray}
	   c_0^2 \rot \rot {\bf E}' =
	 - {\ddot{\bf E}}'
	 + \frac{\partial } { \partial t'} \left( {\bf v} \times \dot{{\bf B}}' \right)
	 \hspace{0mm}\ \nonumber \\
	 - \frac{\partial } { \partial t'}
				\left[ {\bf v}  \left( \nabla {\bf E}' \right)
	           - {\bf E}' \left( \nabla {\bf v}  \right)
				\right]
	 \hspace{0mm}\ \label{B2}
\end{eqnarray}

In order to eliminate the vector of the magnetic field \({\bf B}'\) in Eq.~(\ref{B2}) we resolve Eq.~(\ref{MAX.IN.MT}) with respect to \(\dot{\bf B}'\)
\begin{equation}
	  \dot{\bf B}' = 
	- \rot {\bf E}'
	- \frac{\partial } { \partial t'} \frac{1 } { c_0^2} {\bf v} \times {\bf E}'
\label{MAX_IN_T_B}
\end{equation}

and plug in the right-hand side of Eq.~(\ref{MAX_IN_T_B}) into Eq.\ (\ref{B2}) in order to get an equation featuring only the vector of the electric field \({\bf E}'\).
\begin{widetext}
\[ 
	   c_0^2 \rot \rot {\bf E}' =
	 - {\ddot{\bf E}}'
	 + \frac{\partial } { \partial t'} \left\{ {\bf v} \times 
	     \left[
							- \rot {\bf E}'
							- \frac{\partial } { \partial t'} \frac{1 } { c_0^2} {\bf v} \times {\bf E}'
	     \right] \right\}
	 - \frac{\partial } { \partial t'}
				\left[ {\bf v}  \left( \nabla {\bf E}' \right)
	           -  {\bf E}' \left( \nabla {\bf v} \right)
				\right]
\] 

\[ 
	   c_0^2 \rot \rot {\bf E}' =
	 - {\ddot{\bf E}}'
	 + \frac{\partial } { \partial t'} \left\{
	            - {\bf v} \times \rot {\bf E}'
							- \frac{\partial } { \partial t'} \frac{1 } { c_0^2}
							  {\bf v} \times \left( {\bf v} \times {\bf E}' \right)
	            \right\}
	 - {\bf v}  \left( \nabla \dot{\bf E}' \right)
	 + \left( \nabla {\bf v} \right) \dot{\bf E}'
\] 
	
\[ 
	   c_0^2 \rot \rot {\bf E}' =
	 - {\ddot{\bf E}}'
	 + \frac{\partial } { \partial t'} \left\{
	            - {\bf v} \times \left[ \nabla \times {\bf E}' \right]
							- \frac{\partial } { \partial t'} \frac {1 } { c_0^2}
							  \left[
							    {\bf v} \left( {\bf vE}' \right)
							  - {\bf E}'\left( {\bf vv}  \right)
	              \right] \right\}
	 - {\bf v}  \left( \nabla \dot{\bf E}' \right)
	 + \left( {\bf v} \nabla  \right) \dot{\bf E}'
\] 

\[ 
	   c_0^2 \nabla \times \left( \nabla \times {\bf E}' \right) =
	 - {\ddot{\bf E}}'
	 +  \left\{
	            - \left[   \nabla       \left( {\bf v} \dot{\bf E}' \right)
	                     - \dot{\bf E}' \left( {\bf v} \nabla       \right)
	              \right]
							- \frac{1 } { c_0^2}
							    {\bf v} \left( {\bf v} \ddot{\bf E}' \right)
						  + \frac{v^2 } { c_0^2}\ddot{\bf E}'
	    \right\}
	 - {\bf v}  \left( \nabla \dot{\bf E}' \right)
	 + \left( {\bf v} \nabla  \right) \dot{\bf E}'
\] 

\[ 
		c_0^2 \left[
			\nabla   \left( \nabla {\bf E}' \right) 
		-	{\bf E}' \left( \nabla  \nabla  \right) 
		        \right] =
		- {\ddot{\bf E}}'
		+ \frac{v^2 } { c_0^2}\ddot{\bf E}'
		+ \left( {\bf v} \nabla \right) \dot{\bf E}' 
		- \frac{1 } { c_0^2}{\bf v} \left( {\bf v} \ddot{\bf E}' \right)
		- \nabla \left( {\bf v} \dot{\bf E}' \right)
		- {\bf v}  \left( \nabla \dot{\bf E}' \right)
		+ \left( {\bf v} \nabla  \right) \dot{\bf E}'
\] 

with \(\nabla \nabla = \Delta\) we get
\[ 
			c_0^2 \nabla \left( \nabla {\bf E}' \right) 
		-	c_0^2 \Delta {\bf E}'
		=
		- \frac{1 } { \gamma^2} {\ddot{\bf E}}'
		+ 2 \left( {\bf v} \nabla \right) \dot{\bf E}' 
		- \frac{1 } { c_0^2}{\bf v} \left( {\bf v} \ddot{\bf E}' \right)
		- \nabla \left( {\bf v} \dot{\bf E}' \right)
		- {\bf v}  \left( \nabla \dot{\bf E}' \right)
\] 
\def\theequation{\arabic{equation}}
\begin{equation} 
		 	c_0^2 \Delta {\bf E}'
		- \frac{1 } { \gamma^2} {\ddot{\bf E}}'
		+ 2 \left( {\bf v} \nabla \right) \dot{\bf E}' 
		=
		  \frac{1 } { c_0^2}{\bf v} \left( {\bf v} \ddot{\bf E}' \right)
		+ \nabla \left( {\bf v} \dot{\bf E}' \right)
		+ {\bf v}  \left( \nabla \dot{\bf E}' \right)
		+ c_0^2 \nabla \left( \nabla {\bf E}' \right)
		\tag{\ref{WAVE.EQ.MT}}
\end{equation}   
\def\theequation{\Alph{section}\arabic{equation}}
\end{widetext}
This is the wave equation of the electric field vector \({\bf E}'\) for a moving frame of reference derived from Maxwell Eqs.~(\ref{MAX.DC.MT}) and (\ref{MAX.IN.MT}). -- A similar procedure yields the wave equation of the magnetic field vector \({\bf B}'\), which is identical to Eq.~(\ref{WAVE.EQ.MT}) except that E's are substituted by B's.

\section{Transport of clocks at real speed\label{APP:MOV:CLKS}}

\begin{figure}
	\centering
		\includegraphics{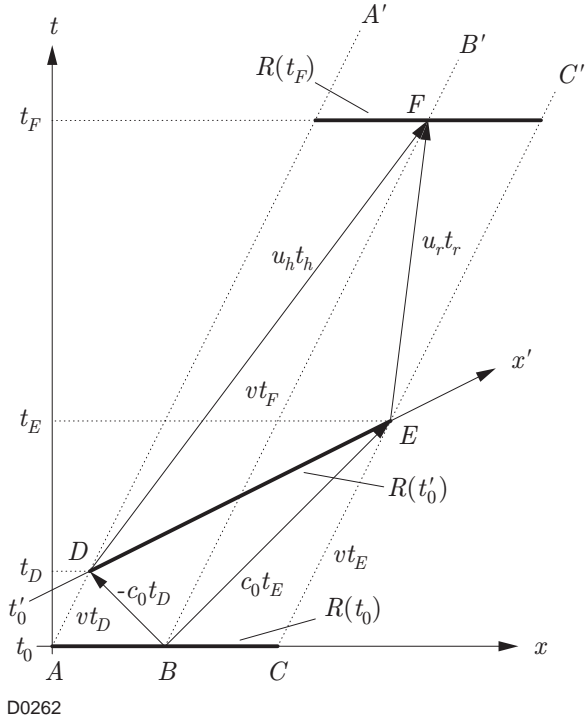}
	\caption{Transport of clocks. Bold lines \(R\) represent a rod of length \(2l_0\) (measured at rest) moving at velocity \(v\) in the positive \(x\)-direction. The rod is shown at three instants \(t_0\), \(t'_0\), and \(t_F\). The dotted lines \(A'\) and \(C'\) are the world-lines of the two endpoints of the rod, dotted line \(B'\) indicates the middle of the rod. \(A\), \(B\), \(C\), \(D\), \(E\), and \(F\) are events in the space-time-like coordinate system \(K_0(x, t)\).---At time \(t_0\) two light pulses \(L_A\) and \(L_B\) are issued from the middle of the rod \(B'\) at event \(B\). When the left-hand light pulse \(L_A\) (arrow \(-c_0 t_D\)) arrives at \(A'\) (event \(D\)), it starts clock \(T_A\) moving towards location \(C'\) at speed \(u_h\). The right-hand light pulse \(L_B\) (arrow \(c_0 t_E\)) starts clock \(T_C\) moving towards \(B'\) at speed \(u_r\) when arriving at \(C'\) (event \(E\)). The bold line from \(D\) to \(E\) represents the `now' of the rod at time \(t'=0\) of the moving system \(K'\). Due to the simultaneously starting light pulses \(L_A\) and \(L_B\), both clocks \(T_A\) and \(T_C\) start simultaneously with respect to the Einstein synchronized moving system \(K'\) and arrive simultaneously in the middle of the rod \(B'\) at event \(F\) when the corresponding transport velocities, measured in the moving system, are of equal magnitude (\(+w\) and \(-w\)). The calculation of the problem shows different elapsed times \(t'_h\) and \(t'_r\) of the clocks \(T_A\) and \(T_C\) when arriving at \(B'\) (although principle of relativity demands equal elapsed times).}
\label{fig:d0262}
\end{figure}

When using a moving clock to synchronize the clocks in an inertial system, the moving clock has to be transported infinitely slowly. At real speed, however, the moving clock no longer transports pure Lorentzian time. We will show this difference by an experiment with two clocks moving at equal speed but in opposite direction with respect to a moving inertial system \(K'\left(x', t'\right)\). The moving system \(K'\) drifts at velocity \(v\) in the positive \(x\)-direction with respect to the system at rest \(K_0\left(x, t\right)\), and is represented by a rod of length \(2\:l_0\) with Einstein synchronized clocks \(T_A\) and \(T_C\) on both ends \(A'\) and \(C'\), thus behaving like a system at rest (see Fig.~\ref{fig:d0262}). The axis of the rod is aligned with the \(x'\)-axis of system \(K'\) as well as with the \(x\)-axis of system \(K_0\) \(\left(x \parallel x'\right)\). Location \(B'\) is exactly in the middle of the rod. -- At \(K_0\)-time \(t_0\) two light pulses \(L_A\) and \(L_B\) are issued from location \(B'\) spreading in the \(-x\)- and \(+x\)-direction, respectively. The light pulses have speed \(c_0\) measured in the stationary system \(K_0\). -- At time \(t_D\) the light pulse \(L_A\) arrives at location \(A'\) and starts clock \(T_A\) moving at speed \(+w\) towards location \(B'\). This event is denoted by \(D\). -- In the opposite direction at time \(t_E\) the light pulse \(L_B\) arrives at location \(C'\) and starts clock \(T_C\) moving towards location \(B'\) at speed \(-w\). This event is denoted by \(E\). -- At time \(t_F\) both clocks \(T_A\) and \(T_C\) meet simultaneously in the middle of the rod at event \(F\).

The velocities \(+w\) and \(-w\) have been measured with clocks and measuring-rods of the moving system \(K'\). The clocks \(T_A\) and \(T_C\) show time zero at the instant when they start moving, thus showing directly elapsed time of movement when arriving at location \(B'\). The principle of relativity demands equal elapsed times of the clocks \(T_A\) and \(T_C\) when arriving in the middle of the rod at event \(F\), but when we calculate this problem from the stationary system \(K_0\), \emph{even with the tools from theory of special relativity}, we obtain a difference in elapsed times, indeed, only a very small, but a nonvanishing difference. The velocities \(+w\) and \(-w\) hold good for the moving system \(K'\). In order to get the corresponding velocities \(u_h\) and \(u_r\) for the system at rest \(K_0\) we have to apply theorem of velocity addition from theory of special relativity, (\ref{UH}) and (\ref{UR}), because of the previously Einstein synchronized clocks \(T_A\) and \(T_C\). They represent Lorentzian time, not medium synchronized time.

We used the vehicle with the light pulses in order to get a simple means for calculation the start events \(D\) and \(E\) without the need for transformation the whole problem into the moving frame. Furthermore, the light pulses ensure simultaneity of the events \(D\) and \(E\) regarded from the moving system \(K'\). In fact, an observer at location \(B'\) will find the clocks starting simultaneously and arriving simultaneously.

First we have to calculate the time coordinates \(t_D\), \(t_E\), and \(t_F\). The light pulses start at event \(B\) when the corresponding time coordinate \(t_0\) equals zero, Eq.~(\ref{T00}).
\begin{equation}
	t_0=0
	\label{T00}
\end{equation}

The distance of events \(A\) and \(C\) is equal to the length of the rod \emph{measured in the system at rest} \(K_0\), thus the distances of events \(A\) and \(B\) as well as of events \(B\) and \(C\) are given by Eq.~(\ref{DIST.AB.BC}).
\begin{equation}
	\overline{AB}=\overline{BC}=l_0 \sqrt{1-\frac{v^2}{c_0^2}}=l_0\frac{1}{\gamma}
	\label{DIST.AB.BC}
\end{equation}

From Eq.~(\ref{PRE.TD}),
\begin{equation}
	-c_0 t_D=-\overline{AB} + v t_D,
	\label{PRE.TD}
\end{equation}
we find the time coordinate \(t_D\) given by Eq.~(\ref{TD}).

\begin{equation}
	t_D=\frac{l_0}{\gamma \left(c_0+v\right)} \label{TD}
\end{equation}

Analogously, we find the time coordinates \(t_E\) and  \(t_F\) being
\begin{equation}
	t_E=\frac{l_0}{\gamma \left(c_0-v\right)} \label{TE}
\end{equation}
and
\begin{equation}
	t_F=\frac{l_0 \gamma}{w} \left(1 + \frac{v}{c_0}\right) \label{TF}.
\end{equation}

The time coordinate \(t_F\) has to be calculated either from triangle (\(BDF\)) as well as from triangle (\(BEF\)) using velocities \(u_h\) and \(u_r\) given by Eqs.~(\ref{UH}) and (\ref{UR}). Both calculations must yield the same result, and indeed they do so, otherwise clocks \(T_A\) and \(T_C\) won't arrive simultaneously at \(B'\) meaning the experiment doesn't work for our purpose.
\begin{equation}
	u_h=\frac{v+w}{1+\frac{vw} {c_0^2}} \label{UH}
\end{equation}

\begin{equation}
	u_r=\frac{v-w}{1-\frac{vw} {c_0^2}} \label{UR}
\end{equation}

The elapsed times \(t_h\) and \(t_r\) for the movment of the clocks \(T_A\) and \(T_C\) measured in the stationary system are simply the differences \(t_F-t_D\) and \(t_F-t_E\), respectively, Eqs.~(\ref{TH}) and (\ref{TR}).
\begin{equation}
	t_h=\frac{l_0 \gamma}{w} \left(1 - 2\:\frac{w}{c_0} + \frac{vw}{c_0^2}\right) \label{TH}
\end{equation}
\begin{equation}
	t_r=\frac{l_0 \gamma}{w} \left(1 - 2\:\frac{w}{c_0} - \frac{vw}{c_0^2}\right) \label{TR}
\end{equation}

In order to achieve elapsed times \(t'_h\) and \(t'_r\) for the clocks \(T_A\) and \(T_C\), i.e., the time the two clocks actually display at event \(F\), we have to take into account time dilation due to the corresponding velocities \(u_h\) and \(u_r\). Using the identities (\ref{ID.H}) and (\ref{ID.R}),
\begin{equation}
	  \sqrt{1-\frac{{u_h}^2}{c_0^2}}
	= \frac{ \sqrt{1-\frac{v^2}{c_0^2}}\sqrt{1-\frac{w^2}{c_0^2}} }
	       {1+\frac{vw}{c_0^2}} 
  \label{ID.H}
\end{equation}

\begin{equation}
	  \sqrt{1-\frac{{u_r}^2}{c_0^2}}
	= \frac{ \sqrt{1-\frac{v^2}{c_0^2}}\sqrt{1-\frac{w^2}{c_0^2}} }
	       {1-\frac{vw}{c_0^2}} ,
  \label{ID.R}
\end{equation}

the calculation of elapsed times \(t'_h\) and \(t'_r\) in the moving frame yields
\begin{equation}
	t'_h = \frac{l_0}{w}
	       \sqrt{1-\frac{w^2}{c_0^2}}
	       \left( 1 - 2\:\frac{w}{c_0}\ \frac{1 - \frac{vw}    {c_0^2} } 
	       																 	 {1 - \frac{v^2w^2}{c_0^4} }
	       \right)
	 \label{TH.PRIME}
\end{equation}

\begin{equation}
  t'_r = \frac{l_0}{w}
	       \sqrt{1-\frac{w^2}{c_0^2}}
	       \left( 1 - 2\:\frac{w}{c_0}\ \frac{1 + \frac{vw}    {c_0^2} } 
	       																	 {1 - \frac{v^2w^2}{c_0^4} }
	       \right).
	 \label{TR.PRIME}
\end{equation}

Finally, the difference \(\Delta t'\) of elapsed times \(t'_h\) and \(t'_r\), calculated from Eqs.~(\ref{TH.PRIME}) and (\ref{TR.PRIME}), yields
\begin{equation}
	\Delta t' = t'_h - t'_r =
	4 \: \frac{l_0 } { c_0} \: \frac{v} {c_0}\: \frac{w} {c_0}\ 
	\frac { \sqrt{1 - \frac{w^2 } { c_0^2 } } }{ 1 - \frac{v^2 w^2 } { c_0^4 } }.
\label{DTD}
\end{equation}

This is the difference of the time readings the two clocks \(T_A\) and \(T_C\) show at the instant they arrive in the middle of the rod at location \(C'\). Please note that despite the remarkable difference \(\Delta t\) of elapsed times \(t_h\) and \(t_r\), measured in the system at rest 
\begin{equation}
	\Delta t=t_h-t_r = 2 \: \frac{l_0}{c_0} \: \frac{v}{c_0} \: \gamma , \label{DELTA.TH.TR}
\end{equation}

the difference \(\Delta t'\) measured with the moving clocks is only a fraction of the order \(2w/c_0\), so we have to say, the moving clocks are really endeavoured to transport Lorentzian time. -- For a drift velocity \(v\) and a transport speed \(w\), both rather less than speed of light \(c_0\), the square root term is approximately unity as well as the corresponding denominator term, and Eq.~(\ref{DTD}) gets the simple form (\ref{DTDS}).
\begin{equation}
	{\begin{array}{c}\hspace{2.0 mm} v \ll c_0\ \\ w \ll c_0 \end{array}} \Bigr\}
	\ \ \Rightarrow \ \ \ 
	\Delta t' \approx 4 \:
  \frac{l_0 } { c_0}\: \frac{v } { c_0}\: \frac{w } { c_0}
\label{DTDS}
\end{equation}

\subsection{Realization of the experiment}

When looking for a possibility to determine the existence of `the medium', the experiment with two moving clocks actually could be one, but regarding the requirements on accuracy and resolution for the measurement of time, distance, and velocity, it is merely a theoretical one as is shown by the result calculated from the following figures: assuming a drift velocity \(v\) of 600 km/s for our galaxy, a transport speed \(w\) for the clocks of 1200 km/h over a distance \(l_0\) of 10 km, and a vacuum speed of light \(c_0\) of 299 792 km/s we have to expect a time difference \(\Delta t'\) in the order of \(3 \times 10^{-13}\) s at a total transport time of 30 s. Besides the enormous challenge for realization of the experimental setup the main problem arises from the requirements on time measurement. We need to have an accuracy for the clocks better than \(10^{-15}\) in order to get reasonable quantities for the time difference. This accuracy is quite beyond the accuracy of currently available clocks suitable for this experiment, meaning that the deviation of the moving time from Lorentzian time isn't measurable yet. There are activities on developing optically controlled clocks aiming on an accuracy of \(10^{-18}\) \cite{Degenhardt-ea.PhysRevA.72.062111}, so far we may expect a possibility in future to get an answer on the question: is there a medium or not? But the same accuracy is recommended for the measurements of velocities \(+w\) and \(-w\), and distances \(\overline{AB}\) and \(\overline{BC}\). The required resolution of distance measurement of its own shows clearly the tremendous difficulties for realization of the experiment. The distances have to be measured with uncertainties in the order of the diameter of an atom, including thermal deviations! Probably, a modification of the experiment eliminates the need for measurement the distance and the velocity at this very high accuracy, and makes it possible to realize the experimental setup. Although the experiment with the two moving clocks cannot be performed at present, the discussion reveals some discrepancy within theory of special relativity, as is shown in the following.

\subsection{Discrepancy within theory of special relativity\label{SAPP:DISCREPANCY}}

The system at rest \(K_0\) may be represented by any inertial system, e.g., the center of the moon, of the sun, or of a cosmic particle rushing through the universe at any speed, or whatever. There are obviously many such systems at rest, and applying principle of relativity, every of these individual Einstein synchronized coordinate systems at rest \(K_{0,i}\) has the same justification \emph{to be} the system at rest \(K_0\). The time differences \(\Delta t'_i\) for each of the individual systems \(K_{0,i}\), calculated from Eq.\ (\ref{DTD}), will depend on the drift velocities \(v_i\) measured from the individual systems. Thus, due to principle of relativity, any number of different time differences \(\Delta t'_i\) have to be expected from the experiment with the moving clocks, \emph{at the same time}\,! Here, we have probably exposed an intrinsic contradiction of theory of special relativity, because the experiment will yield only one single value \(\Delta t'_m\) for the time difference, without any doubt. This \(\Delta t'_m\) matches solely with the \(\Delta t'_p\) of one particular system  \(K_{0,p}\) from all systems at rest \(K_{0,i}\), and this reference system \(K_{0,p}\) can be regarded as something like a \emph{preferred} inertial system, the system really at rest. All the other Einstein synchronized inertial systems \(K_{0,i \not = p}\) only behave like systems at rest, but actually they are moving systems, only virtually at rest. Probably, our universe isn't that relative as it appears to be, and it looks like the concept of a preferred frame, outcast by the principle of relativity, has crept into special relativity through the backdoor.

We shall mention that medium transformation has no problem with a preferred frame of reference. Applying medium transformation on the above transport experiment yields only one single value for the time difference \(\Delta t'\), holding good for all reference frames involved.

\section{Remarks on the medium\label{APP.MED}}

Initially, I had no intention to give any statement on a proposed medium, because there is no theory of a medium at all, only a few ideas exist, therefore any statement is rather speculative. And here, I don't even ask for the reason to introduce a medium, I ask only for the presupposition. Along with the discussions accompaning the formation of this contribution the questions often occurred: What is the medium alike? What is the nature of this medium? -- Therefore, I felt it is indicated at least to characterize briefly the current ideas to a model of the medium, however, keeping in mind these are only ideas, which are subject to change whenever it is required by future development.

Our medium has an elastic structure with a homogenious mass density when no distortion occures. The elastic properties are characterized by a kind of a stress tensor with components independent of the position when no distortion is present. The structure is able to carry repulsive forces, attractive forces, as well as torsional forces, and other kinds of elastic forces. Thus there are many kinds of elastic waves possible, given by the mass density and the stress tensor. The energy density of these elastic waves obeys \(1/r^{2}\)-law. Light quanta are a special kind of waves, maybe a kind of a torsional wave, occurring as wave packets, which don't spread like a sphere but along a straight line, thus conserving the amplitude and the energy of the packets (at least on a small scale). Particles are frozen packets of that torsional waves getting stable by a special kind of a torsion state of the surrounding medium, which is equivalent to the charge of the particle \footnote{
There are actually two particles, which are known to be stable, the electron and the proton (and of course the corresponding anti-particle), both particles are `charged'.}. The frozen wave packets as well as the free running light quanta cause a distortion of mass density in such a manner, that within the space occupied by the wave packet the mass density is enhanced, and the mass density in the vicinity outside the wave packet is decreased. The components of the stress tensor are also altered, and the deviation of these components from the equilibrium value at no distortion as well as the deviation of the mass density decay according to \(1/r\)-law thus forming the gravitational field of a particle or of a light quantum. The particle or the frozen wave packet, respectively, isn't fixed at a certain position within the medium, but can be moved througout the medium at any speed, provided this speed is less than the speed of the free running wave packets, i.e., the speed of light \(c_0\).

An ensemble of particles forms a body. The medium drifts through moving particles and through moving bodies without remarkable friction, regardless to the number of particles comprising that body. This holds good for small bodies as well as for very large bodies such as the moon, the earth, and even for the sun. A body represents a system of reference, and the speed of light is only isotropic in that system when the drift velocity of this body is zero with respect to the medium. When the drift velocity is different from zero, we have anisotropy of speed of light. This is a fundamental property of our proposed medium, which we have to take into account when looking for a transformation of coordinates from a system at rest into a moving system. Furthermore our model makes clear that speed of light, measured in a particular system of reference, is independent from the velocity of the emitter. Of course, the measured magnitude surely depends on the moving state of this system, say, it depends on the synchronization scheme applied to the clocks of the system of reference.

\bibliography{weiss}

\end{document}